\documentclass[aps,prl,twocolumn,superscriptaddress]{revtex4-2}
\pdfoutput=1

\usepackage{amsmath,amsfonts,amssymb,amsthm}
\usepackage{graphicx}
\usepackage{xcolor}
\usepackage{scalerel}
\usepackage[colorlinks=true,linkcolor=black,citecolor=black,urlcolor=blue,bookmarks,breaklinks=true]{hyperref}

\newcommand\figBST{S1}
\newcommand\figDE{S2}
\newcommand\figDphi{S3}
\newcommand\figSize{S4}

\newcommand\videoNetworks{S1}
\newcommand\videoBST{S2}

\newcommand{\nc}{\newcommand}
\nc{\QWS}{\scaleto{\text{QWS}}{5pt}}
\nc{\SQ}{\mathcal{S}_{\QWS}(\alpha)}
\renewcommand\vec{\mathbf}

\begin{document}

\title{%
Gap Sensitivity Reveals Universal Behaviors in Optimized Photonic \texorpdfstring{\\}{}
Crystal and Disordered Networks
}

\author{Michael A. Klatt}
\email{mklatt@princeton.edu}
\affiliation{Department of Physics, Princeton University, Princeton, New Jersey 08544, USA}
\affiliation{Institut für Theoretische Physik, University of Erlangen-Nürnberg, Staudtstr. 7, 91058 Erlangen, Germany}
\author{Paul J. Steinhardt} 
\email{steinh@princeton.edu}
\affiliation{Department of Physics, Princeton University, Princeton, New Jersey 08544, USA}
\author{Salvatore Torquato}
\email{torquato@princeton.edu}
\affiliation{Department of Physics, Princeton University, Princeton, New Jersey 08544, USA}
\affiliation{Department of Chemistry, Princeton Institute for the 
 	     Science and Technology of Materials, and Program in Applied and 
 	     Computational Mathematics, Princeton University, Princeton, New Jersey 
 	     08544, USA}
\date{\today}

\begin{abstract}
\noindent  %
Through an extensive series of high-precision numerical computations of 
the optimal complete photonic band gap (PBG) as a function of dielectric 
contrast $\alpha$ for a variety of crystal and disordered 
heterostructures, we reveal striking universal behaviors of the 
\textit{gap sensitivity} $\mathcal{S}(\alpha)\equiv 
d\Delta(\alpha)/d\alpha$, the first derivative of the optimal 
gap-to-midgap ratio $\Delta(\alpha)$.
In particular, for all our crystal networks, $\mathcal{S}(\alpha)$ takes 
a universal form that is  well approximated by the analytic formula for 
a one-dimensional quarter-wave stack, $\SQ$.
Even more surprisingly, the values of $\mathcal{S}(\alpha)$ for our disordered 
networks converge to $\SQ$ for sufficiently large $\alpha$.
A deeper understanding of the simplicity of this universal behavior may 
provide fundamental insights about PBG formation and guidance in the 
design of novel photonic heterostructures.
\end{abstract}


\maketitle

\begin{figure}[b]
  \centering %
  \includegraphics[width=\linewidth]{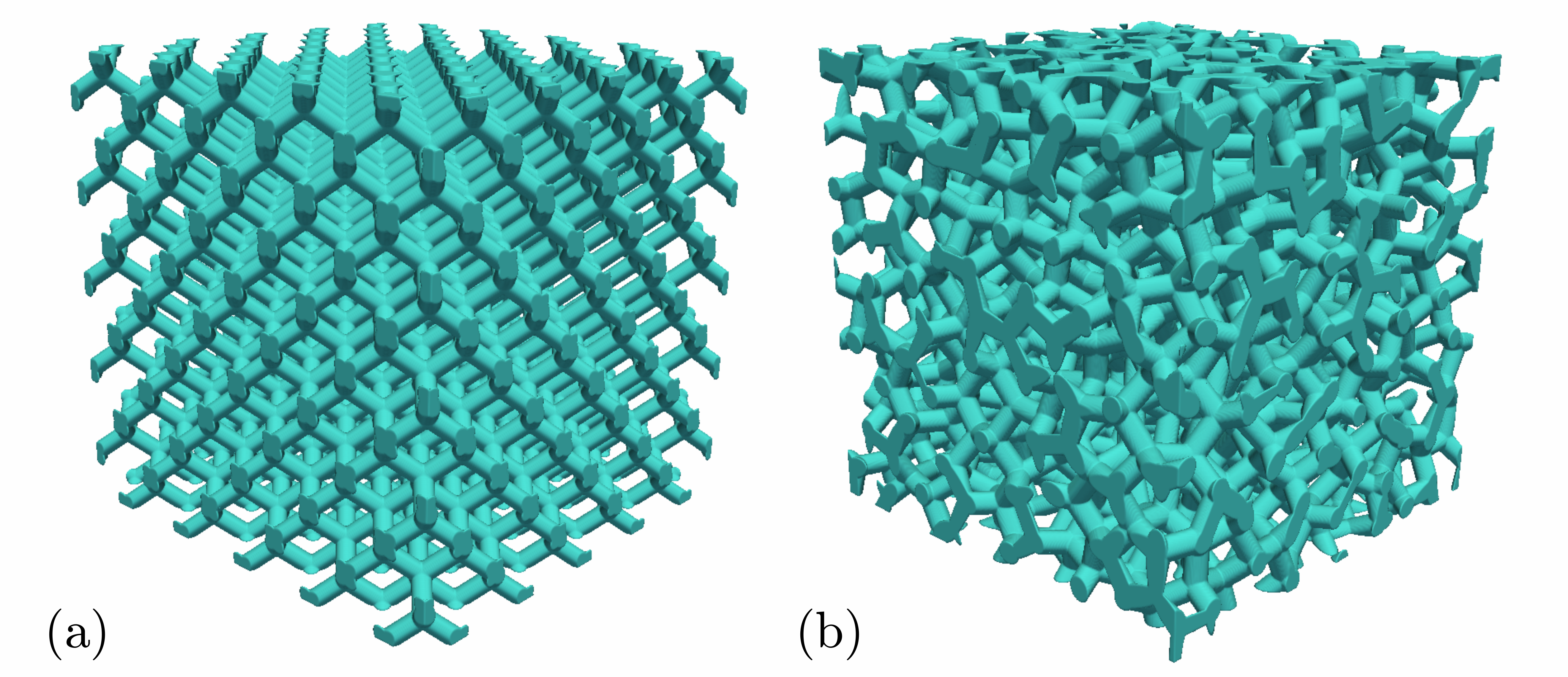}
  \caption{Photonic networks: (a) the crystal diamond network, where 
    cylindrical rods connect nearest neighbors of a diamond lattice, and 
    (b) the disordered nearly hyperuniform network (NHN) model; see also 
    video~\videoNetworks{}.
    \label{fig:samples}}
\end{figure}

\paragraph{Introduction.}
A complete photonic band gap (PBG) prohibits the propagation of light in 
all directions and for all polarizations for a substantial continuous 
range of frequencies~\cite{joannopoulos_photonic_2008, 
maldovan_diamond-structured_2004}.
PBGs can occur in heterostructures composed of two or more substances 
with different dielectric constants such as silicon and air; see 
Fig.~\ref{fig:samples}.
In early studies, PBGs were exclusively found in crystalline structures, 
such as the diamond crystal network~\cite{chan_photonic_1991}.
Later it was discovered that quasicrystals without long-range periodic 
translational order~\cite{man_experimental_2005, 
rechtsman_optimized_2008} and isotropic ``disordered'' solids can also 
exhibit complete PBGs~\cite{florescu_designer_2009, 
edagawa_photonic_2008, imagawa_photonic_2010, liew_photonic_2011, 
yin_amorphous_2012, edagawa_photonic_2014, muller_silicon_2014, 
tsitrin_unfolding_2015, imagawa_robustness_2016, sellers_local_2017, 
li_biological_2018}.

In this paper, we present evidence for a subtle, unanticipated universal 
behavior of the maximum complete PBG as a function of dielectric contrast 
among network heterostructures spanning a wide variety of symmetries and 
topologies.
For this purpose, we follow a multistage procedure that begins with 
identifying different ``candidate classes'' of networks distinguished by 
their nearest-neighbor table properties (\textit{e.g., coordination}), 
translational order and rotational symmetry which are known from past 
work to include examples with large complete 
PBGs~\cite{joannopoulos_photonic_2008, maldovan_diamond-structured_2004, 
chan_photonic_1991, florescu_designer_2009, edagawa_photonic_2008, 
men_robust_2014}.
For each candidate class, we aim to identify the member with the 
largest complete PBG, which are, based on experience to date, among 
those that combine the highest degree of 
hyperuniformity~\cite{torquato_hyperuniform_2018, hejna_nearly_2013, 
deringer_realistic_2018, klatt_universal_2019} with sufficiently narrow 
distributions of bond lengths and angles.

Then, we optimize the microscopic properties of the selected networks.
For this, we assume the networks are composed of rods and spheres 
with dielectric constant $\varepsilon_2$ embedded in a bulk that has 
a smaller dielectric constant $\varepsilon_1$.
We also assume the rods have circular cross-section with radius $R$ 
joined at sphere-shaped vertices with the same radius $R$.
For each fixed dielectric contrast ratio 
$\alpha:=\varepsilon_2/\varepsilon_1$, we vary $R$ to find the 
\textit{maximal value of the photonic band gap-to-midgap ratio} 
$\Delta(\alpha)$; that is, %
\begin{align}
  \Delta(\alpha) := \max_{R\geq 0}
  \left\{\frac{\Delta\omega}{\omega_m}\big(R,\alpha\big)\right\}.
  \label{eq:delta}
\end{align}
We call the optimized radius $R_{\mathrm{opt}}(\alpha)$ and the 
corresponding optimized volume fraction $\phi_{\mathrm{opt}}(\alpha)$.

Since our purpose is to study cases with large PBGs, we only consider 
candidate classes known to have some networks satisfying this condition.
In some cases, this condition is not straightforward to check.
For example, disordered networks typically have many localized defect 
modes that break up what would otherwise be a large PBG into many 
smaller PBGs.
As a last step of the optimization, we check whether such defects can be 
removed via bond switching; if so, then the candidate class is included 
in the study and the modified network is treated according to the same 
optimization procedure as described above.

Critical to our analysis are high-precision calculations of the optimal 
gap-to-midgap ratios, $\Delta(\alpha)$, that are unprecedented in their 
scope, both for the wide variety of heterostructures considered and the wide 
range of dielectric contrasts.
Performing this computation for disordered networks with 1000 
vertices is only first possible now using state-of-the-art computational 
techniques and computer clusters.
To get reliable results for $\Delta(\alpha)$ and hence 
$\mathcal{S}(\alpha)$, it is essential to (i) accurately determine the 
optimized volume fraction $\phi_{\mathrm{opt}}(\alpha)$; (ii) compute the 
stopgap along a number of different directions of propagation; (iii) use a 
sufficiently high spatial resolution; and (iv) use a precise plane wave 
expansion method to solve Maxwell's 
equation~\cite{johnson_block-iterative_2001,CPUTime}.

A cursory view of plots of $\Delta(\alpha)$ as a function of $\alpha$ 
already suggests certain common trends across the different types of 
networks (despite the distinctly different functional values).
What proves to be the key to revealing the universal behaviors is the 
{\it gap-sensitivity}, $\mathcal{S}(\alpha)$, defined as
\begin{align}
  \mathcal{S}(\alpha) := \frac{d\Delta(\alpha)}{d \alpha}.
  \label{eq:slope}
\end{align}
As we show below, for all our optimized three-dimensional {\it crystal} networks, 
$\mathcal{S}(\alpha)$ is well approximated for all $\alpha$ by %
\begin{align}
  \SQ:=\frac{2}{\pi}\frac{1}{\alpha^{3/4}(1+\sqrt{\alpha})}, 
  \label{eq:qws}
\end{align}
where $\SQ$ is the precise analytic result for the gap-sensitivity of a 
one-dimensional \textit{quarter-wave stack} composed of alternating layers with 
dielectric constants $\varepsilon_1$ and $\varepsilon_2$ and each of 
quarter-wavelength thickness (which results in an optimized volume 
fraction) where the propagation of electromagnetic waves is 
perpendicular to the alternating dielectric layers.
For a derivation of the corresponding formula for $\Delta(\alpha)$; see, 
e.g., Refs.~\cite{yeh_optical_1988, macleod_thin-film_2010}.

Perhaps even more striking is the fact that, for all our optimized {\it 
disordered} networks, the values of $\mathcal{S}(\alpha)$ {\it also converge to} $\SQ$ 
at large $\alpha$.
In other words, at sufficiently large $\alpha$, all our optimized 
crystal and disordered heterostructures~---~despite the differences in 
symmetry, topology, and long-range order~---~exhibit the same universal 
behavior: $\mathcal{S}(\alpha) \approx \SQ$.

Notably, the story is different for small $\alpha = {\cal O}(1)$ near 
the critical value $\alpha_c$ (defined as the minimal contrast at which 
a complete PBG first appears).
Over this range, $\mathcal{S}(\alpha)$ displays a common pattern for all 
our optimized disordered networks that is clearly distinct from the 
common pattern for crystal networks.
Namely, whereas $\mathcal{S}(\alpha)$ for the {\it crystal} networks 
decreases monotonically even for small $\alpha$, $\mathcal{S}(\alpha)$ 
for {\it disordered} networks first increases, reaches a maximum, and 
then decreases, ultimately converging to $\SQ$ as $\alpha$ continues to 
increase.

\paragraph{Candidate classes of networks.}
Among the abundant variety of known photonic crystals and disordered 
heterostructures, we select, for our computationally intensive 
optimization, candidate classes of networks that represent a broad 
spectrum of symmetries and coordination numbers.
% Crystals 
The diamond crystal network exhibits the largest known optimal 
gap-to-midgap ratio $\Delta(\alpha)$~\cite{chan_photonic_1991, 
maldovan_diamond-structured_2004, men_robust_2014}.
The hexagonal-diamond network has the same topology and perfectly 
tetrahedral vertices but a different symmetry.
It has substantial PBGs, although its optimal gap-to-midgap ratios 
$\Delta(\alpha)$ are distinctly smaller than those of the diamond 
network~\cite{klatt_phoamtonic_2019}.
A better performing photonic crystal is 
the rod-connected network based on the Laves graph, which has the same 
topology as the single gyroid~\cite{maldovan_diamond-structured_2004, 
hyde_short_2008, sellers_local_2017, wilts_butterfly_2017}.
The diamond and Laves graphs are the only two crystal lattices in three 
dimensions with the ``strong isotropic property''~\cite{sunada_crystals_2008}, 
i.e., a symmetry under permutations of neighboring edges.
The key differences are that the diamond network is fourfold 
coordinated and nonchiral with a face-centered cubic symmetry and the 
Laves graph is trivalent and chiral with a body-centered cubic symmetry.
We also study a simple cubic (SC) heterostructure, a sixfold-coordinated 
network consisting of rods connecting nearest neighbors in a SC 
lattice~\cite{sozuer_photonic_1993, maldovan_photonic_2005}.

% Disordered networks 
Our isotropic disordered networks are not based on any underlying 
lattice.
They exhibit a correlated disorder with a varying degree of both local 
and global order.
Based on experience to date, the member in a candidate class with the 
largest PBG has bond length and angle distributions with a standard 
deviation less than 15\% of the mean.
For larger variations, the PBG becomes smaller and the absolute 
difference between $\mathcal{S}(\alpha)$ and $\SQ$ becomes larger.

One class of disordered models that we consider are continuous random
networks (CRNs), i.e., idealized models for amorphous tetrahedrally
coordinated solids (like amorphous
silicon)~\cite{barkema_high-quality_2000, hejna_nearly_2013}.
The first complete PBG of three-dimensional disordered networks was
found for CRNs~\cite{edagawa_photonic_2008}.
Here, we study the PBGs of the nearly hyperuniform network (NHN)
model~\cite{hejna_nearly_2013}.
Starting from a classical CRN, the model is carefully annealed
to suppress large-scale density fluctuations.
The model is thus driven towards a vanishing of density fluctuations
in the infinite-wavelength limit, known as
hyperuniformity~\cite{torquato_local_2003, torquato_hyperuniform_2018}.
In two dimensions, hyperuniformity has been found advantageous for
opening up large complete PBGs~\cite{florescu_designer_2009}.

Another disordered network that we consider is based on an alternative 
structural model of amorphous silicon that was simulated by a slow 
quench of a liquid using molecular dynamics 
(MD)~\cite{deringer_realistic_2018}.
When we connected each atom with its four nearest neighbors, we obtained 
a disordered photonic network, but it had defect modes that 
appeared within a large PBG.
We, therefore, switched bonds to remove the defect modes, nearly 
doubling the gap size.

Finally, we introduce quantizer-based networks (QBNs).
We begin from amorphous, nearly hyperuniform inherent structures of the 
quantizer energy starting from some random initial point 
configuration~\cite{klatt_universal_2019}.
For more details on the quantizer energy functional; see supplemental 
material (SM) Sec.~1 and Refs.~\cite{torquato_reformulation_2010, 
lloyd_least_1982, liu_centroidal_2009, du_advances_2010, zhang_periodic_2012, 
ruscher_voronoi_2015, ruscher_glassy_2020, hain_low-temperature_2020, 
klatt_universal_2019}.
Next, we construct the corresponding Delaunay tessellation.
A QBN then connects the centroids of neighboring cells, in analogy to 
the tiling procedure for hyperuniform disordered solids (HUDS) in two 
dimensions described in Ref.~\cite{florescu_designer_2009}.
In three dimensions, the tiling procedure was already applied to hard-sphere 
packings~\cite{liew_photonic_2011, haberko_fabrication_2013, 
muller_silicon_2014}
and vertex models~\cite{li_biological_2018}.
To avoid defect modes for $\alpha > 13$, we removed each vertex where 
two triangles met by bond switching.

Our three disordered networks fulfill all of our selection criteria.
In particular, the ratio of standard deviations to mean values for the 
bond lengths and angles are 
 3.6\% and  7.4\% for the NHN,
 1.9\% and  8.5\% for the MD quench, and
 7.4\% and 10.7\% for the QBN.

\begin{figure}[t]
  \centering \includegraphics[width=\linewidth]{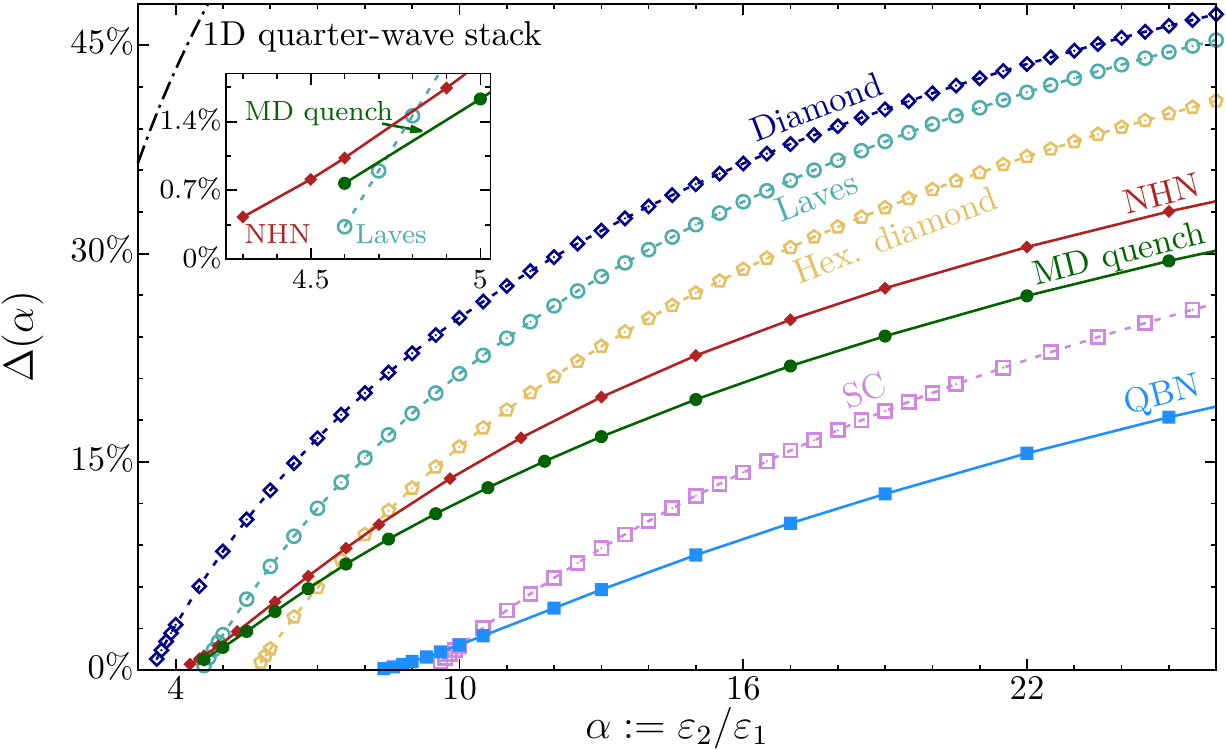}
  \caption{The \textit{gap plot} shows the optimal gap-to-midgap ratio 
    $\Delta(\alpha)$ as a function of the dielectric contrast $\alpha$.
    The plot compares three-dimensional photonic crystals (open symbols) to three-dimensional 
    disordered networks (solid symbols).
    The dashed-dotted line shows $\Delta_{\QWS}(\alpha)$ for the 
    one-dimensional 
    quarter-wave stack.
    The inset zooms in on small values of $\alpha$ where the PBGs first 
    open up for the NHN, MD quench, and Laves networks.
    \label{fig:gap}}
\end{figure}

\paragraph{Protocol.}
All samples are constructed with periodic boundary conditions to enable 
band structure calculations.
The PBGs are accurately determined using the plane-wave expansion method 
to solve the frequency-domain eigenproblem implemented in the MIT 
Photonic Bands (\textsc{MPB}) software 
package~\cite{johnson_block-iterative_2001}.
The plane-wave expansion is the best available method for achieving the 
high precision needed for the optimization and analysis presented here.
It is not clear whether the same accuracy can be achieved using 
time-domain eigensolvers, like the finite-difference time-domain (FDTD) 
method, which may miss eigenfrequencies or locate spurious 
ones~\cite{joannopoulos_photonic_2008}.
We define a complete PBG by a range of frequencies for which there are 
no states no matter the direction of the wave vector.
In practice, we compute the stopgaps for a finite number of directions 
along the edges of the irreducible Brillouin zone.

For each sample and each dielectric contrast $\alpha$, we optimize the 
rod radius $R$ and hence the volume fraction $\phi$ of the high 
dielectric phase.
We first determine the gap-to-midgap ratios $\Delta\omega/\omega_m$ for 
several radii close to the putative optimum PBG.
Next, we fit a parabola to estimate the optimized radius 
$R_{\mathrm{opt}}(\alpha)$, at which we then repeat the band structure 
calculations.
Thus, we determine the \textit{optimal} gap-to-midgap ratio 
$\Delta(\alpha)$; see Eq.~\eqref{eq:delta}.
We estimate $\mathcal{S}(\alpha)$, see Eq.~\eqref{eq:slope}, 
using the symmetric difference quotient.

The combined absolute error depends on the type of network and 
dielectric contrast and is significantly smaller at low than at high 
$\alpha$.
Therefore, we limit our investigation to values of $\alpha \leq 26$.
The error in determining $\Delta(\alpha)$ is less than 0.7\% for $\alpha \leq 13$ 
and less than about 1.4\% for $\alpha > 13$.
As shown in Fig.~\figSize{}, the results are not sensitive to system 
size.

\begin{figure}[t]
  \centering 
  \includegraphics[width=\linewidth]{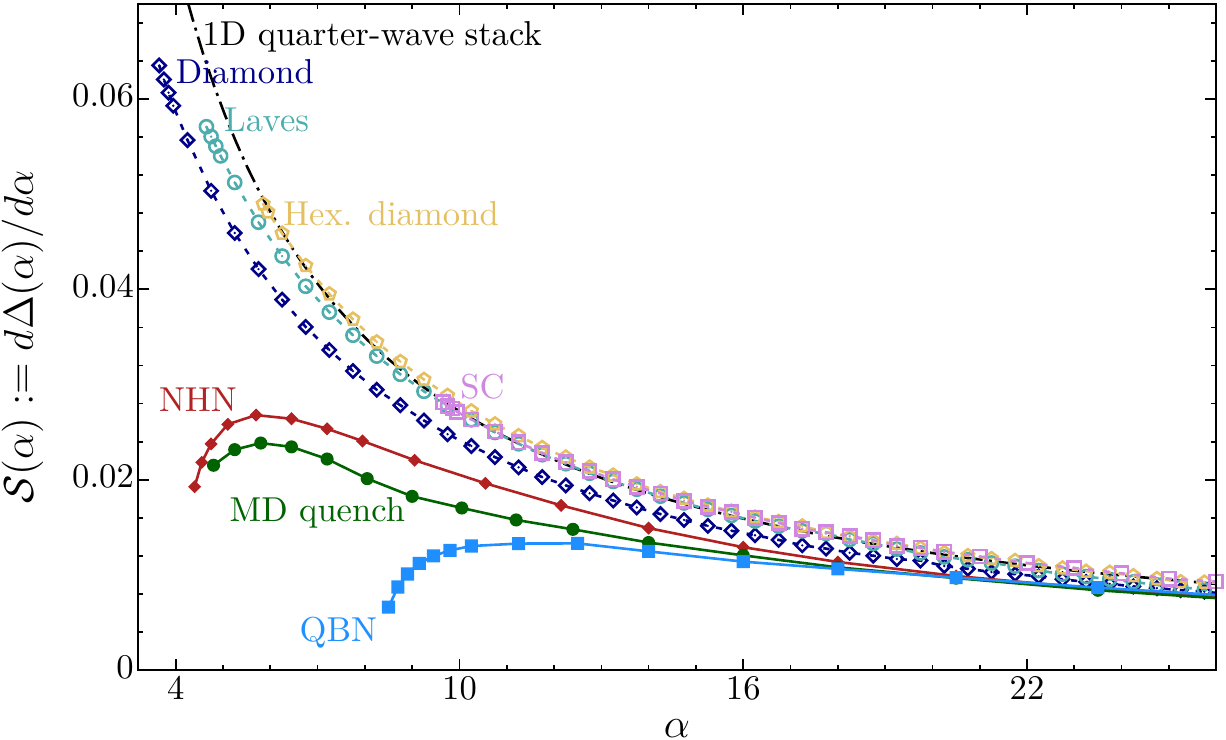}
  \caption{The \textit{gap-sensitivity plot} shows the 
  slope $\mathcal{S}(\alpha):=d\Delta(\alpha)/d\alpha$ for each of the 
  curves in Fig.~\ref{fig:gap}.
  For our three-dimensional crystal networks, $\mathcal{S}(\alpha)$ is well 
  approximated by $\SQ$ for all $\alpha$.
  For our disordered networks, $\mathcal{S}(\alpha)$ has a maximum at 
  low $\alpha = {\cal O}(1)$ and only converges (approximately) to $\SQ$
  at larger $\alpha \gg 1$.
  \label{fig:gapsensitivity}}
\end{figure}

\paragraph{Gap plots and gap-sensitivity plots.}
The key results  of this paper derive from Figs.~\ref{fig:gap} 
and~\ref{fig:gapsensitivity}.
Figure~\ref{fig:gap} shows plots of the optimal gap-to-midgap ratio 
$\Delta(\alpha)$ as a function of $\alpha$ (referred to henceforth as 
the  \textit{gap plot}) for all of our crystal and disordered 
networks with different symmetries and topological characteristics.
Figure~\ref{fig:gapsensitivity} (the \textit{gap-sensitivity plot}) 
shows the slope $\mathcal{S}(\alpha):=d\Delta(\alpha)/d\alpha$ for 
the same networks.

\paragraph{Results.}
Photonic band structures of optimized crystal and disordered 
networks considered in this work are shown in Fig.~\figBST{} and 
video~\videoBST{}.
The gap plot in Fig.~\ref{fig:gap} shows that the diamond network 
exhibits the largest PBG for all dielectric contrasts, closely followed 
by the Laves network.
At $\alpha > 10$, disordered networks have smaller PBGs than the Laves 
and hexagonal-diamond networks.
However, $\Delta(\alpha)$ of the disordered networks increases more 
slowly as a function of $\alpha$ than those of the crystal networks, 
leading to a notable crossing of curves in the gap plot.
In fact, we find a smaller critical contrast, $\alpha_c=4.1$, for the 
NHN model and MD quench than for the hexagonal-diamond network 
($\alpha_{c}=5.8$) or even the Laves network ($\alpha_{c}=4.5$).
Only the diamond network has a smaller critical contrast 
($\alpha_{c}=3.5$).
Moreover, the gap plot shows that the NHN model has a larger optimal 
gap-to-midgap ratio [$\Delta(13)=19.7\%$] than the currently reported 
record for CRNs (18\%) at $\alpha=13$~\cite{edagawa_photonic_2008}.

Turning now to the gap-sensitivity plot in 
Fig.~\ref{fig:gapsensitivity}, we note that for all our crystal networks 
and for all $\alpha$, $\mathcal{S}(\alpha)$ is well approximated without 
any fit parameter by $\SQ$ from Eq.~\eqref{eq:qws}, the analytic result 
for the quarter-wave stack.
Remarkably, even for our isotropic disordered networks, 
$\mathcal{S}(\alpha)$ converges at large $\alpha$ to $\SQ$ to a good 
approximation.
More precisely, we use ``convergence'' in this paper to mean 
$\mathcal{S}(\alpha) = (1 \pm 0.2) \SQ$ for $\alpha \gtrsim 20$,
corresponding to an absolute error of $2\times 10^{-3}$,
which we estimate to be the systematic error associated with the 
optimization protocol used to identify the member of 
a candidate class with the topology, bond length and angle 
distributions, and volume fraction that maximizes the PBG.
Convergence to within this 20\% uncertainty is impressive given the 
diverse candidate classes that have been considered: classes with highly 
isotropic and anisotropic PBGs; with uniform and irregular topologies; 
with translational order and disorder; and with one-dimensional layered structure and 
three-dimensional network structure.

We emphasize that the multistage optimization protocol is essential to 
achieving this level of convergence.  Without any optimization, the 
difference between $\mathcal{S}(\alpha)$ and $\SQ$ would increase by an 
order of magnitude or more.
Even though $\mathcal{S}(\alpha)$ is approximately the same for our 
three, four, and sixfold coordinated networks, we only find the 
universal behavior for networks with relatively uniform bond lengths and 
angles, as required in the first stage of our procedure.
To examine the effect of variations in bond lengths and angles, we 
randomly perturbed the vertices of a diamond network using independent 
and isotropic Gaussian displacements.
The strength of perturbations is controlled by the ratio $a$ of the 
standard deviation of the displacements to the bond length in the 
unperturbed network.
For $a \leq 10\%$, the perturbed diamond network fulfills our selection 
criterion for bond length and angles, as mentioned above, and 
$\mathcal{S}(\alpha)$ approximately converges to $\SQ$ at large 
$\alpha$.
However, with increasing strength of perturbations $a$, we find that the 
difference between $\mathcal{S}(\alpha)$ and $\SQ$ increasingly grows 
until the PBG closes altogether when $a\approx 0.5$.
We find the same behavior for a perturbed 
CRN~\cite{barkema_high-quality_2000}.
The qualitative behavior of $\mathcal{S}(\alpha)$ remains the same, but 
$\mathcal{S}(\alpha)$ deviates from $\SQ$ approximately by a constant 
factor that grows as $a$ increases.

The second stage of our procedure entails finding the optimized 
volume fraction.
This step is crucial because $\phi_{\mathrm{opt}}(\alpha)$ varies 
between 8.0\% and 38.4\% for our networks and values of $\alpha$ so 
that, without optimizing $\phi$, we obtain both quantitatively and 
qualitatively different results.
If we only optimized $\phi$ to within a certain percentage of the 
true optimal value, we observe similar effects as for large variations 
in the bond lengths and angles: the difference between 
$\mathcal{S}(\alpha)$ and $\SQ$ increases with an increasing percentage difference 
between $\phi$ and $\phi_{\mathrm{opt}}(\alpha)$.
Alternatively, if we fix $\phi$ independent of $\alpha$, we observe a 
more dramatic effect:
the gap-to-midgap ratio quickly flattens for large $\alpha$ at a value 
well below the optimal $\Delta(\alpha)$ of that candidate class; 
that is, $\mathcal{S}(\alpha)$ falls rapidly to zero at large 
$\alpha$.

Finally, the third stage is necessary to take account of localized defect modes 
that break up an otherwise large PBG.
These defects not only decrease $\Delta(\alpha)$ but also lead to large 
deviations between $\mathcal{S}(\alpha)$ and $\SQ$.
We observed such localized defects in the initial QBN model 
wherever two triangles met at one vertex.
When we removed these triangle defects by bond switching, 
$\Delta(\alpha)$ increased significantly and $\mathcal{S}(\alpha)$ 
approached $\SQ$ at large $\alpha$.

Although we do not have a theoretical explanation of the universal 
behavior of $\mathcal{S}(\alpha)$ at large $\alpha$ and the systematic 
differences between crystal versus disordered networks at small 
$\alpha$, we find an interesting correlation with the behavior of the 
square of the magnitude of the electric field eigenmodes 
$\vec{E}(\vec{r},\alpha)$ just below and above the PBG.
More precisely, we find that $\|\vec{E}(\vec{r},\alpha)\|^2$ changes 
more at small $\alpha$ for a given change in dielectric contrast 
$\alpha$ than it does at large $\alpha$ and that this effect is 
much more pronounced for the disordered than for the crystal networks.
To quantify this effect, we introduce, as a heuristic measure, the 
average change of $\|\vec{E}(\vec{r},\alpha)\|^2$ with the dielectric contrast:
$\mathcal{D}_{\vec{E}}(\alpha):=\int\left|{\partial\|\vec{E}(\vec{r},\alpha)\|^2}/{\partial\alpha}\right|d\vec{r}$.
Figure~\figDE{} shows that $\mathcal{D}_{\vec{E}}(\alpha)$ 
is, apart from statistical and numerical fluctuations, a decreasing 
function.
At low $\alpha$, the changes are distinctly stronger for the disordered 
than the crystal networks.
In contrast, at large $\alpha$, where we observe the universal behavior 
of $\mathcal{S}(\alpha)$, $\mathcal{D}_{\vec{E}}(\alpha)$ 
converges within the computational uncertainty for the crystal and 
disordered networks.

This observation suggests the following argument:
At large $\alpha$, the electric field is strongly confined by the 
high-dielectric material, and the configuration changes only slowly with 
the dielectric contrast.
At small $\alpha$, the electric field is generically less confined to 
the high-dielectric material allowing more degrees of freedom to be 
accessed when optimizing the PBG.
The high degree of translational order in crystal systems constrains the 
field configuration possibilities, so the corresponding curves of 
$\mathcal{S}(\alpha)$ still follow Eq.~\eqref{eq:qws}.
For disordered systems, though, there are not the same symmetry 
constraints, allowing more field configurations that enable the 
optimized PBG to decrease more slowly as $\alpha$ decreases.

We also observe that the derivative of the optimal volume fraction, 
$d\phi_{\mathrm{opt}}(\alpha)/d\alpha$, approximately agrees for all our 
three-dimensional crystal networks at all $\alpha$ and is, in fact, well approximated 
by the analytic result for the one-dimensional quarter-wave stack: 
$d\phi_{\QWS}(\alpha)/d\alpha=-1/(2\alpha^{3/2}+4\alpha+2\sqrt{\alpha})$.
Similarly, for our disordered networks, 
$d\phi_{\mathrm{opt}}(\alpha)/d\alpha$ approximately converges to the 
same formula at large $\alpha$; see Fig.~\figDphi{}.
This behavior correlates with our finding that the optimization of the volume 
fraction is an essential part of our multistage procedure to reveal the 
universal behavior of $\mathcal{S}(\alpha)$.

\paragraph{Conclusions and Outlook.}
For all our candidate classes, including ordered and disordered varieties 
with different symmetries and topologies, $\mathcal{S}(\alpha)$ of our 
optimized three-dimensional networks approximately converges at large $\alpha$ to 
$\SQ$ of the one-dimensional quarter-wave stack.
A physical explanation of this universal behavior has to apply to both 
crystal and disordered networks even though the PBGs in these two cases 
have distinctly different physical properties.
The crystal networks have anisotropic PBGs bounded above at one 
$k$ point and below by a different $k$ point along the band gap.
The slope $\mathcal{S}(\alpha)$ of the single stop gaps does not agree 
as well with the one-dimensional quarter-wave stack formula as that of the 
three-dimensional 
complete PBG.
In contrast, the PBGs of the disordered networks are isotropic, i.e., 
they have the same stop gap in every direction (apart from negligible 
statistical fluctuations) so that the upper and lower boundaries of 
the PBG are both set by any single $k$ point.
Moreover, the eigenmodes above and below the PBGs of the disordered 
networks are localized~---~again in contrast to the crystal networks.
Finally, a physical explanation has to explicate why it only 
applies if the networks are optimized according to the three stages of 
our procedure.

We believe the similar behaviors of $\mathcal{S}(\alpha)$, 
$\mathcal{D}_{\vec{E}}(\alpha)$, and 
$\phi_{\mathrm{opt}}(\alpha)$ provide important clues for understanding 
the universal curves of $\mathcal{S}(\alpha)$ reported here, but we have 
not yet found a solid theoretical explanation that ties these different 
observations together to explain the universal behaviors.
The challenge is to identify how, from a combination of highly 
nonlinear physics and a diverse range of network geometries and 
topologies explored here, these behaviors emerge.

As a practical application, the discovery of this universal behavior 
at large $\alpha$ makes it possible to estimate for a given type of 
optimized heterostructure the photonic band gap-to-midgap ratio 
$\Delta(\alpha)$ for all $\alpha$ once one has determined it for a 
single $\alpha$ without any further extensive computations.
On the more theoretical side, the fact that the behavior applies to 
optimized structures in three dimensions, whether crystalline or disordered, as well 
as to periodic layered media, indicates an unanticipated simplicity, 
despite the apparent nonlinear mathematics and complex physics involved 
in the PBG computation and the optimization procedure.
A better understanding of this behavior will ultimately help to identify 
the relation between the structural features of optimized 
heterostructures and the formation of large PBGs, which may guide the 
design of improved photonic heterostructures.

\begin{acknowledgments}
We thank G.\,T.~Barkema and N.~Mousseau for providing samples of CRNs and 
the Princeton Institute for Computational Science and Engineering 
(PICSciE) for the computational resources.
This work was partially supported by the Princeton University Innovation Fund for 
New Ideas in the Natural Sciences.
S.\,T.~also gratefully acknowledges the support of the Air Force Office of 
Scientific Research Program on Mechanics of Multifunctional Materials 
and Microsystems under Grant No.~FA9550-18-1-0514.
M.\,A.\,K.~also acknowledges funding by the Volkswagenstiftung via the 
Experiment-Projekt Mecke.
\end{acknowledgments}

\bibliography{gap-sensitivity}

\begin{thebibliography}{42}%
\makeatletter
\providecommand \@ifxundefined [1]{%
 \@ifx{#1\undefined}
}%
\providecommand \@ifnum [1]{%
 \ifnum #1\expandafter \@firstoftwo
 \else \expandafter \@secondoftwo
 \fi
}%
\providecommand \@ifx [1]{%
 \ifx #1\expandafter \@firstoftwo
 \else \expandafter \@secondoftwo
 \fi
}%
\providecommand \natexlab [1]{#1}%
\providecommand \enquote  [1]{``#1''}%
\providecommand \bibnamefont  [1]{#1}%
\providecommand \bibfnamefont [1]{#1}%
\providecommand \citenamefont [1]{#1}%
\providecommand \href@noop [0]{\@secondoftwo}%
\providecommand \href [0]{\begingroup \@sanitize@url \@href}%
\providecommand \@href[1]{\@@startlink{#1}\@@href}%
\providecommand \@@href[1]{\endgroup#1\@@endlink}%
\providecommand \@sanitize@url [0]{\catcode `\\12\catcode `\$12\catcode
  `\&12\catcode `\#12\catcode `\^12\catcode `\_12\catcode `\%12\relax}%
\providecommand \@@startlink[1]{}%
\providecommand \@@endlink[0]{}%
\providecommand \url  [0]{\begingroup\@sanitize@url \@url }%
\providecommand \@url [1]{\endgroup\@href {#1}{\urlprefix }}%
\providecommand \urlprefix  [0]{URL }%
\providecommand \Eprint [0]{\href }%
\providecommand \doibase [0]{https://doi.org/}%
\providecommand \selectlanguage [0]{\@gobble}%
\providecommand \bibinfo  [0]{\@secondoftwo}%
\providecommand \bibfield  [0]{\@secondoftwo}%
\providecommand \translation [1]{[#1]}%
\providecommand \BibitemOpen [0]{}%
\providecommand \bibitemStop [0]{}%
\providecommand \bibitemNoStop [0]{.\EOS\space}%
\providecommand \EOS [0]{\spacefactor3000\relax}%
\providecommand \BibitemShut  [1]{\csname bibitem#1\endcsname}%
\let\auto@bib@innerbib\@empty
%</preamble>
\bibitem [{\citenamefont {Joannopoulos}\ \emph {et~al.}(2008)\citenamefont
  {Joannopoulos}, \citenamefont {Johnson}, \citenamefont {Winn},\ and\
  \citenamefont {Meade}}]{joannopoulos_photonic_2008}%
  \BibitemOpen
  \bibfield  {author} {\bibinfo {author} {\bibfnamefont {J.~D.}\ \bibnamefont
  {Joannopoulos}}, \bibinfo {author} {\bibfnamefont {S.~G.}\ \bibnamefont
  {Johnson}}, \bibinfo {author} {\bibfnamefont {J.~N.}\ \bibnamefont {Winn}},\
  and\ \bibinfo {author} {\bibfnamefont {R.~D.}\ \bibnamefont {Meade}},\
  }\href@noop {} {\emph {\bibinfo {title} {Photonic Crystals: Molding the Flow
  of Light}}},\ \bibinfo {edition} {2nd}\ ed.\ (\bibinfo  {publisher}
  {{Princeton University Press}},\ \bibinfo {address} {{Princeton}},\ \bibinfo
  {year} {2008})\BibitemShut {NoStop}%
\bibitem [{\citenamefont {Maldovan}\ and\ \citenamefont
  {Thomas}(2004)}]{maldovan_diamond-structured_2004}%
  \BibitemOpen
  \bibfield  {author} {\bibinfo {author} {\bibfnamefont {M.}~\bibnamefont
  {Maldovan}}\ and\ \bibinfo {author} {\bibfnamefont {E.~L.}\ \bibnamefont
  {Thomas}},\ }\href {https://doi.org/10.1038/nmat1201} {\bibfield  {journal}
  {\bibinfo  {journal} {Nat. Mater.}\ }\textbf {\bibinfo {volume} {3}},\
  \bibinfo {pages} {593} (\bibinfo {year} {2004})}\BibitemShut {NoStop}%
\bibitem [{\citenamefont {Chan}\ \emph {et~al.}(1991)\citenamefont {Chan},
  \citenamefont {Ho},\ and\ \citenamefont {Soukoulis}}]{chan_photonic_1991}%
  \BibitemOpen
  \bibfield  {author} {\bibinfo {author} {\bibfnamefont {C.~T.}\ \bibnamefont
  {Chan}}, \bibinfo {author} {\bibfnamefont {K.~M.}\ \bibnamefont {Ho}},\ and\
  \bibinfo {author} {\bibfnamefont {C.~M.}\ \bibnamefont {Soukoulis}},\ }\href
  {https://doi.org/10.1209/0295-5075/16/6/009} {\bibfield  {journal} {\bibinfo
  {journal} {Europhys. Lett.}\ }\textbf {\bibinfo {volume} {16}},\ \bibinfo
  {pages} {563} (\bibinfo {year} {1991})}\BibitemShut {NoStop}%
\bibitem [{\citenamefont {Man}\ \emph {et~al.}(2005)\citenamefont {Man},
  \citenamefont {Megens}, \citenamefont {Steinhardt},\ and\ \citenamefont
  {Chaikin}}]{man_experimental_2005}%
  \BibitemOpen
  \bibfield  {author} {\bibinfo {author} {\bibfnamefont {W.}~\bibnamefont
  {Man}}, \bibinfo {author} {\bibfnamefont {M.}~\bibnamefont {Megens}},
  \bibinfo {author} {\bibfnamefont {P.~J.}\ \bibnamefont {Steinhardt}},\ and\
  \bibinfo {author} {\bibfnamefont {P.~M.}\ \bibnamefont {Chaikin}},\ }\href
  {https://doi.org/10.1038/nature03977} {\bibfield  {journal} {\bibinfo
  {journal} {Nature}\ }\textbf {\bibinfo {volume} {436}},\ \bibinfo {pages}
  {993} (\bibinfo {year} {2005})}\BibitemShut {NoStop}%
\bibitem [{\citenamefont {Rechtsman}\ \emph {et~al.}(2008)\citenamefont
  {Rechtsman}, \citenamefont {Jeong}, \citenamefont {Chaikin}, \citenamefont
  {Torquato},\ and\ \citenamefont {Steinhardt}}]{rechtsman_optimized_2008}%
  \BibitemOpen
  \bibfield  {author} {\bibinfo {author} {\bibfnamefont {M.~C.}\ \bibnamefont
  {Rechtsman}}, \bibinfo {author} {\bibfnamefont {H.-C.}\ \bibnamefont
  {Jeong}}, \bibinfo {author} {\bibfnamefont {P.~M.}\ \bibnamefont {Chaikin}},
  \bibinfo {author} {\bibfnamefont {S.}~\bibnamefont {Torquato}},\ and\
  \bibinfo {author} {\bibfnamefont {P.~J.}\ \bibnamefont {Steinhardt}},\ }\href
  {https://doi.org/10.1103/PhysRevLett.101.073902} {\bibfield  {journal}
  {\bibinfo  {journal} {Phys. Rev. Lett.}\ }\textbf {\bibinfo {volume} {101}},\
  \bibinfo {pages} {073902} (\bibinfo {year} {2008})}\BibitemShut {NoStop}%
\bibitem [{\citenamefont {Florescu}\ \emph {et~al.}(2009)\citenamefont
  {Florescu}, \citenamefont {Torquato},\ and\ \citenamefont
  {Steinhardt}}]{florescu_designer_2009}%
  \BibitemOpen
  \bibfield  {author} {\bibinfo {author} {\bibfnamefont {M.}~\bibnamefont
  {Florescu}}, \bibinfo {author} {\bibfnamefont {S.}~\bibnamefont {Torquato}},\
  and\ \bibinfo {author} {\bibfnamefont {P.~J.}\ \bibnamefont {Steinhardt}},\
  }\href {https://doi.org/10.1073/pnas.0907744106} {\bibfield  {journal}
  {\bibinfo  {journal} {Proc. Natl. Acad. Sci. U.S.A.}\ }\textbf {\bibinfo
  {volume} {106}},\ \bibinfo {pages} {20658} (\bibinfo {year}
  {2009})}\BibitemShut {NoStop}%
\bibitem [{\citenamefont {Edagawa}\ \emph {et~al.}(2008)\citenamefont
  {Edagawa}, \citenamefont {Kanoko},\ and\ \citenamefont
  {Notomi}}]{edagawa_photonic_2008}%
  \BibitemOpen
  \bibfield  {author} {\bibinfo {author} {\bibfnamefont {K.}~\bibnamefont
  {Edagawa}}, \bibinfo {author} {\bibfnamefont {S.}~\bibnamefont {Kanoko}},\
  and\ \bibinfo {author} {\bibfnamefont {M.}~\bibnamefont {Notomi}},\ }\href
  {https://doi.org/10.1103/PhysRevLett.100.013901} {\bibfield  {journal}
  {\bibinfo  {journal} {Phys. Rev. Lett.}\ }\textbf {\bibinfo {volume} {100}},\
  \bibinfo {pages} {013901} (\bibinfo {year} {2008})}\BibitemShut {NoStop}%
\bibitem [{\citenamefont {Imagawa}\ \emph {et~al.}(2010)\citenamefont
  {Imagawa}, \citenamefont {Edagawa}, \citenamefont {Morita}, \citenamefont
  {Niino}, \citenamefont {Kagawa},\ and\ \citenamefont
  {Notomi}}]{imagawa_photonic_2010}%
  \BibitemOpen
  \bibfield  {author} {\bibinfo {author} {\bibfnamefont {S.}~\bibnamefont
  {Imagawa}}, \bibinfo {author} {\bibfnamefont {K.}~\bibnamefont {Edagawa}},
  \bibinfo {author} {\bibfnamefont {K.}~\bibnamefont {Morita}}, \bibinfo
  {author} {\bibfnamefont {T.}~\bibnamefont {Niino}}, \bibinfo {author}
  {\bibfnamefont {Y.}~\bibnamefont {Kagawa}},\ and\ \bibinfo {author}
  {\bibfnamefont {M.}~\bibnamefont {Notomi}},\ }\href
  {https://doi.org/10.1103/PhysRevB.82.115116} {\bibfield  {journal} {\bibinfo
  {journal} {Phys. Rev. B}\ }\textbf {\bibinfo {volume} {82}},\ \bibinfo
  {pages} {115116} (\bibinfo {year} {2010})}\BibitemShut {NoStop}%
\bibitem [{\citenamefont {Liew}\ \emph {et~al.}(2011)\citenamefont {Liew},
  \citenamefont {Yang}, \citenamefont {Noh}, \citenamefont {Schreck},
  \citenamefont {Dufresne}, \citenamefont {O'Hern},\ and\ \citenamefont
  {Cao}}]{liew_photonic_2011}%
  \BibitemOpen
  \bibfield  {author} {\bibinfo {author} {\bibfnamefont {S.~F.}\ \bibnamefont
  {Liew}}, \bibinfo {author} {\bibfnamefont {J.-K.}\ \bibnamefont {Yang}},
  \bibinfo {author} {\bibfnamefont {H.}~\bibnamefont {Noh}}, \bibinfo {author}
  {\bibfnamefont {C.~F.}\ \bibnamefont {Schreck}}, \bibinfo {author}
  {\bibfnamefont {E.~R.}\ \bibnamefont {Dufresne}}, \bibinfo {author}
  {\bibfnamefont {C.~S.}\ \bibnamefont {O'Hern}},\ and\ \bibinfo {author}
  {\bibfnamefont {H.}~\bibnamefont {Cao}},\ }\href
  {https://doi.org/10.1103/PhysRevA.84.063818} {\bibfield  {journal} {\bibinfo
  {journal} {Phys. Rev. A}\ }\textbf {\bibinfo {volume} {84}},\ \bibinfo
  {pages} {063818} (\bibinfo {year} {2011})}\BibitemShut {NoStop}%
\bibitem [{\citenamefont {Yin}\ \emph {et~al.}(2012)\citenamefont {Yin},
  \citenamefont {Dong}, \citenamefont {Liu}, \citenamefont {Zhan},
  \citenamefont {Shi}, \citenamefont {Zi},\ and\ \citenamefont
  {Yablonovitch}}]{yin_amorphous_2012}%
  \BibitemOpen
  \bibfield  {author} {\bibinfo {author} {\bibfnamefont {H.}~\bibnamefont
  {Yin}}, \bibinfo {author} {\bibfnamefont {B.}~\bibnamefont {Dong}}, \bibinfo
  {author} {\bibfnamefont {X.}~\bibnamefont {Liu}}, \bibinfo {author}
  {\bibfnamefont {T.}~\bibnamefont {Zhan}}, \bibinfo {author} {\bibfnamefont
  {L.}~\bibnamefont {Shi}}, \bibinfo {author} {\bibfnamefont {J.}~\bibnamefont
  {Zi}},\ and\ \bibinfo {author} {\bibfnamefont {E.}~\bibnamefont
  {Yablonovitch}},\ }\href {https://doi.org/10.1073/pnas.1204383109} {\bibfield
   {journal} {\bibinfo  {journal} {Proc. Natl. Acad. Sci. U.S.A.}\ }\textbf
  {\bibinfo {volume} {109}},\ \bibinfo {pages} {10798} (\bibinfo {year}
  {2012})}\BibitemShut {NoStop}%
\bibitem [{\citenamefont {Edagawa}(2014)}]{edagawa_photonic_2014}%
  \BibitemOpen
  \bibfield  {author} {\bibinfo {author} {\bibfnamefont {K.}~\bibnamefont
  {Edagawa}},\ }\href {https://doi.org/10.1088/1468-6996/15/3/034805}
  {\bibfield  {journal} {\bibinfo  {journal} {Sci. Technol. Adv. Mater.}\
  }\textbf {\bibinfo {volume} {15}},\ \bibinfo {pages} {034805} (\bibinfo
  {year} {2014})}\BibitemShut {NoStop}%
\bibitem [{\citenamefont {Muller}\ \emph {et~al.}(2014)\citenamefont {Muller},
  \citenamefont {Haberko}, \citenamefont {Marichy},\ and\ \citenamefont
  {Scheffold}}]{muller_silicon_2014}%
  \BibitemOpen
  \bibfield  {author} {\bibinfo {author} {\bibfnamefont {N.}~\bibnamefont
  {Muller}}, \bibinfo {author} {\bibfnamefont {J.}~\bibnamefont {Haberko}},
  \bibinfo {author} {\bibfnamefont {C.}~\bibnamefont {Marichy}},\ and\ \bibinfo
  {author} {\bibfnamefont {F.}~\bibnamefont {Scheffold}},\ }\href
  {https://doi.org/10.1002/adom.201300415} {\bibfield  {journal} {\bibinfo
  {journal} {Adv. Opt. Mater.}\ }\textbf {\bibinfo {volume} {2}},\ \bibinfo
  {pages} {115} (\bibinfo {year} {2014})}\BibitemShut {NoStop}%
\bibitem [{\citenamefont {Tsitrin}\ \emph {et~al.}(2015)\citenamefont
  {Tsitrin}, \citenamefont {Williamson}, \citenamefont {Amoah}, \citenamefont
  {Nahal}, \citenamefont {Chan}, \citenamefont {Florescu},\ and\ \citenamefont
  {Man}}]{tsitrin_unfolding_2015}%
  \BibitemOpen
  \bibfield  {author} {\bibinfo {author} {\bibfnamefont {S.}~\bibnamefont
  {Tsitrin}}, \bibinfo {author} {\bibfnamefont {E.~P.}\ \bibnamefont
  {Williamson}}, \bibinfo {author} {\bibfnamefont {T.}~\bibnamefont {Amoah}},
  \bibinfo {author} {\bibfnamefont {G.}~\bibnamefont {Nahal}}, \bibinfo
  {author} {\bibfnamefont {H.~L.}\ \bibnamefont {Chan}}, \bibinfo {author}
  {\bibfnamefont {M.}~\bibnamefont {Florescu}},\ and\ \bibinfo {author}
  {\bibfnamefont {W.}~\bibnamefont {Man}},\ }\href
  {https://doi.org/10.1038/srep13301} {\bibfield  {journal} {\bibinfo
  {journal} {Sci. Rep.}\ }\textbf {\bibinfo {volume} {5}},\ \bibinfo {pages}
  {13301} (\bibinfo {year} {2015})}\BibitemShut {NoStop}%
\bibitem [{\citenamefont {Imagawa}\ and\ \citenamefont
  {Edagawa}(2017)}]{imagawa_robustness_2016}%
  \BibitemOpen
  \bibfield  {author} {\bibinfo {author} {\bibfnamefont {S.}~\bibnamefont
  {Imagawa}}\ and\ \bibinfo {author} {\bibfnamefont {K.}~\bibnamefont
  {Edagawa}},\ }\href {https://doi.org/10.1007/s00339-016-0703-6} {\bibfield
  {journal} {\bibinfo  {journal} {Appl. Phys. A}\ }\textbf {\bibinfo {volume}
  {123}},\ \bibinfo {pages} {41} (\bibinfo {year} {2017})}\BibitemShut
  {NoStop}%
\bibitem [{\citenamefont {Sellers}\ \emph {et~al.}(2017)\citenamefont
  {Sellers}, \citenamefont {Man}, \citenamefont {Sahba},\ and\ \citenamefont
  {Florescu}}]{sellers_local_2017}%
  \BibitemOpen
  \bibfield  {author} {\bibinfo {author} {\bibfnamefont {S.~R.}\ \bibnamefont
  {Sellers}}, \bibinfo {author} {\bibfnamefont {W.}~\bibnamefont {Man}},
  \bibinfo {author} {\bibfnamefont {S.}~\bibnamefont {Sahba}},\ and\ \bibinfo
  {author} {\bibfnamefont {M.}~\bibnamefont {Florescu}},\ }\href
  {https://doi.org/10.1038/ncomms14439} {\bibfield  {journal} {\bibinfo
  {journal} {Nat. Commun.}\ }\textbf {\bibinfo {volume} {8}},\ \bibinfo {pages}
  {14439} (\bibinfo {year} {2017})}\BibitemShut {NoStop}%
\bibitem [{\citenamefont {Li}\ \emph {et~al.}(2018)\citenamefont {Li},
  \citenamefont {Das},\ and\ \citenamefont {Bi}}]{li_biological_2018}%
  \BibitemOpen
  \bibfield  {author} {\bibinfo {author} {\bibfnamefont {X.}~\bibnamefont
  {Li}}, \bibinfo {author} {\bibfnamefont {A.}~\bibnamefont {Das}},\ and\
  \bibinfo {author} {\bibfnamefont {D.}~\bibnamefont {Bi}},\ }\href
  {https://doi.org/10.1073/pnas.1715810115} {\bibfield  {journal} {\bibinfo
  {journal} {Proc. Natl. Acad. Sci. U.S.A.}\ }\textbf {\bibinfo {volume}
  {115}},\ \bibinfo {pages} {6650} (\bibinfo {year} {2018})}\BibitemShut
  {NoStop}%
\bibitem [{\citenamefont {Men}\ \emph {et~al.}(2014)\citenamefont {Men},
  \citenamefont {Lee}, \citenamefont {Freund}, \citenamefont {Peraire},\ and\
  \citenamefont {Johnson}}]{men_robust_2014}%
  \BibitemOpen
  \bibfield  {author} {\bibinfo {author} {\bibfnamefont {H.}~\bibnamefont
  {Men}}, \bibinfo {author} {\bibfnamefont {K.~Y.~K.}\ \bibnamefont {Lee}},
  \bibinfo {author} {\bibfnamefont {R.~M.}\ \bibnamefont {Freund}}, \bibinfo
  {author} {\bibfnamefont {J.}~\bibnamefont {Peraire}},\ and\ \bibinfo {author}
  {\bibfnamefont {S.~G.}\ \bibnamefont {Johnson}},\ }\href
  {https://doi.org/10.1364/OE.22.022632} {\bibfield  {journal} {\bibinfo
  {journal} {Opt. Express}\ }\textbf {\bibinfo {volume} {22}},\ \bibinfo
  {pages} {22632} (\bibinfo {year} {2014})}\BibitemShut {NoStop}%
\bibitem [{\citenamefont {Torquato}(2018)}]{torquato_hyperuniform_2018}%
  \BibitemOpen
  \bibfield  {author} {\bibinfo {author} {\bibfnamefont {S.}~\bibnamefont
  {Torquato}},\ }\href {https://doi.org/10.1016/j.physrep.2018.03.001}
  {\bibfield  {journal} {\bibinfo  {journal} {Phys. Rep.}\ }\textbf {\bibinfo
  {volume} {745}},\ \bibinfo {pages} {1} (\bibinfo {year} {2018})}\BibitemShut
  {NoStop}%
\bibitem [{\citenamefont {Hejna}\ \emph {et~al.}(2013)\citenamefont {Hejna},
  \citenamefont {Steinhardt},\ and\ \citenamefont
  {Torquato}}]{hejna_nearly_2013}%
  \BibitemOpen
  \bibfield  {author} {\bibinfo {author} {\bibfnamefont {M.}~\bibnamefont
  {Hejna}}, \bibinfo {author} {\bibfnamefont {P.~J.}\ \bibnamefont
  {Steinhardt}},\ and\ \bibinfo {author} {\bibfnamefont {S.}~\bibnamefont
  {Torquato}},\ }\href {https://doi.org/10.1103/PhysRevB.87.245204} {\bibfield
  {journal} {\bibinfo  {journal} {Phys. Rev. B}\ }\textbf {\bibinfo {volume}
  {87}},\ \bibinfo {pages} {245204} (\bibinfo {year} {2013})}\BibitemShut
  {NoStop}%
\bibitem [{\citenamefont {Deringer}\ \emph {et~al.}(2018)\citenamefont
  {Deringer}, \citenamefont {Bernstein}, \citenamefont {Bart{\'o}k},
  \citenamefont {Cliffe}, \citenamefont {Kerber}, \citenamefont {Marbella},
  \citenamefont {Grey}, \citenamefont {Elliott},\ and\ \citenamefont
  {Cs{\'a}nyi}}]{deringer_realistic_2018}%
  \BibitemOpen
  \bibfield  {author} {\bibinfo {author} {\bibfnamefont {V.~L.}\ \bibnamefont
  {Deringer}}, \bibinfo {author} {\bibfnamefont {N.}~\bibnamefont {Bernstein}},
  \bibinfo {author} {\bibfnamefont {A.~P.}\ \bibnamefont {Bart{\'o}k}},
  \bibinfo {author} {\bibfnamefont {M.~J.}\ \bibnamefont {Cliffe}}, \bibinfo
  {author} {\bibfnamefont {R.~N.}\ \bibnamefont {Kerber}}, \bibinfo {author}
  {\bibfnamefont {L.~E.}\ \bibnamefont {Marbella}}, \bibinfo {author}
  {\bibfnamefont {C.~P.}\ \bibnamefont {Grey}}, \bibinfo {author}
  {\bibfnamefont {S.~R.}\ \bibnamefont {Elliott}},\ and\ \bibinfo {author}
  {\bibfnamefont {G.}~\bibnamefont {Cs{\'a}nyi}},\ }\href
  {https://doi.org/10.1021/acs.jpclett.8b00902} {\bibfield  {journal} {\bibinfo
   {journal} {J. Phys. Chem. Lett.}\ }\textbf {\bibinfo {volume} {9}},\
  \bibinfo {pages} {2879} (\bibinfo {year} {2018})}\BibitemShut {NoStop}%
\bibitem [{\citenamefont {Klatt}\ \emph
  {et~al.}(2019{\natexlab{a}})\citenamefont {Klatt}, \citenamefont
  {Lovri{\'c}}, \citenamefont {Chen}, \citenamefont {Kapfer}, \citenamefont
  {Schaller}, \citenamefont {Sch{\"o}nh{\"o}fer}, \citenamefont {Gardiner},
  \citenamefont {Smith}, \citenamefont {{Schr{\"o}der-Turk}},\ and\
  \citenamefont {Torquato}}]{klatt_universal_2019}%
  \BibitemOpen
  \bibfield  {author} {\bibinfo {author} {\bibfnamefont {M.~A.}\ \bibnamefont
  {Klatt}}, \bibinfo {author} {\bibfnamefont {J.}~\bibnamefont {Lovri{\'c}}},
  \bibinfo {author} {\bibfnamefont {D.}~\bibnamefont {Chen}}, \bibinfo {author}
  {\bibfnamefont {S.~C.}\ \bibnamefont {Kapfer}}, \bibinfo {author}
  {\bibfnamefont {F.~M.}\ \bibnamefont {Schaller}}, \bibinfo {author}
  {\bibfnamefont {P.~W.~A.}\ \bibnamefont {Sch{\"o}nh{\"o}fer}}, \bibinfo
  {author} {\bibfnamefont {B.~S.}\ \bibnamefont {Gardiner}}, \bibinfo {author}
  {\bibfnamefont {A.-S.}\ \bibnamefont {Smith}}, \bibinfo {author}
  {\bibfnamefont {G.~E.}\ \bibnamefont {{Schr{\"o}der-Turk}}},\ and\ \bibinfo
  {author} {\bibfnamefont {S.}~\bibnamefont {Torquato}},\ }\href
  {https://doi.org/10.1038/s41467-019-08360-5} {\bibfield  {journal} {\bibinfo
  {journal} {Nat. Commun.}\ }\textbf {\bibinfo {volume} {10}},\ \bibinfo
  {pages} {811} (\bibinfo {year} {2019}{\natexlab{a}})}\BibitemShut {NoStop}%
\bibitem [{\citenamefont {Johnson}\ and\ \citenamefont
  {Joannopoulos}(2001)}]{johnson_block-iterative_2001}%
  \BibitemOpen
  \bibfield  {author} {\bibinfo {author} {\bibfnamefont {S.}~\bibnamefont
  {Johnson}}\ and\ \bibinfo {author} {\bibfnamefont {J.}~\bibnamefont
  {Joannopoulos}},\ }\href {https://doi.org/10.1364/OE.8.000173} {\bibfield
  {journal} {\bibinfo  {journal} {Opt. Express}\ }\textbf {\bibinfo {volume}
  {8}},\ \bibinfo {pages} {173} (\bibinfo {year} {2001})}\BibitemShut {NoStop}%
\bibitem [{CPUTime()}]{CPUTime}%
  \BibitemOpen
  \bibinfo {note} {{The overall CPU time utilized for this study was about
  70~years on Intel Skylake CPUs.}}\BibitemShut {Stop}%
\bibitem [{\citenamefont {Yeh}(1988)}]{yeh_optical_1988}%
  \BibitemOpen
  \bibfield  {author} {\bibinfo {author} {\bibfnamefont {P.}~\bibnamefont
  {Yeh}},\ }\href@noop {} {\emph {\bibinfo {title} {Optical Waves in Layered
  Media}}}\ (\bibinfo  {publisher} {{Wiley}},\ \bibinfo {address} {{Hoboken,
  NJ}},\ \bibinfo {year} {1988})\BibitemShut {NoStop}%
\bibitem [{\citenamefont {Macleod}(2010)}]{macleod_thin-film_2010}%
  \BibitemOpen
  \bibfield  {author} {\bibinfo {author} {\bibfnamefont {H.~A.}\ \bibnamefont
  {Macleod}},\ }\href {https://doi.org/10.1887/0750306882} {\emph {\bibinfo
  {title} {Thin-{{Film Optical Filters}}}}},\ \bibinfo {edition} {4th}\ ed.\
  (\bibinfo  {publisher} {{CRC Press}},\ \bibinfo {address} {{Boca Raton;
  London; New York}},\ \bibinfo {year} {2010})\BibitemShut {NoStop}%
\bibitem [{\citenamefont {Klatt}\ \emph
  {et~al.}(2019{\natexlab{b}})\citenamefont {Klatt}, \citenamefont
  {Steinhardt},\ and\ \citenamefont {Torquato}}]{klatt_phoamtonic_2019}%
  \BibitemOpen
  \bibfield  {author} {\bibinfo {author} {\bibfnamefont {M.~A.}\ \bibnamefont
  {Klatt}}, \bibinfo {author} {\bibfnamefont {P.~J.}\ \bibnamefont
  {Steinhardt}},\ and\ \bibinfo {author} {\bibfnamefont {S.}~\bibnamefont
  {Torquato}},\ }\href {https://doi.org/10.1073/pnas.1912730116} {\bibfield
  {journal} {\bibinfo  {journal} {Proc. Natl. Acad. Sci. U.S.A.}\ }\textbf
  {\bibinfo {volume} {116}},\ \bibinfo {pages} {23480} (\bibinfo {year}
  {2019}{\natexlab{b}})}\BibitemShut {NoStop}%
\bibitem [{\citenamefont {Hyde}\ \emph {et~al.}(2008)\citenamefont {Hyde},
  \citenamefont {O'Keeffe},\ and\ \citenamefont {Proserpio}}]{hyde_short_2008}%
  \BibitemOpen
  \bibfield  {author} {\bibinfo {author} {\bibfnamefont {S.~T.}\ \bibnamefont
  {Hyde}}, \bibinfo {author} {\bibfnamefont {M.}~\bibnamefont {O'Keeffe}},\
  and\ \bibinfo {author} {\bibfnamefont {D.~M.}\ \bibnamefont {Proserpio}},\
  }\href {https://doi.org/10.1002/anie.200801519} {\bibfield  {journal}
  {\bibinfo  {journal} {Angew. Chem. Int. Ed.}\ }\textbf {\bibinfo {volume}
  {47}},\ \bibinfo {pages} {7996} (\bibinfo {year} {2008})}\BibitemShut
  {NoStop}%
\bibitem [{\citenamefont {Wilts}\ \emph {et~al.}(2017)\citenamefont {Wilts},
  \citenamefont {Apeleo~Zubiri}, \citenamefont {Klatt}, \citenamefont {Butz},
  \citenamefont {Fischer}, \citenamefont {Kelly}, \citenamefont {Spiecker},
  \citenamefont {Steiner},\ and\ \citenamefont
  {{Schr{\"o}der-Turk}}}]{wilts_butterfly_2017}%
  \BibitemOpen
  \bibfield  {author} {\bibinfo {author} {\bibfnamefont {B.~D.}\ \bibnamefont
  {Wilts}}, \bibinfo {author} {\bibfnamefont {B.}~\bibnamefont
  {Apeleo~Zubiri}}, \bibinfo {author} {\bibfnamefont {M.~A.}\ \bibnamefont
  {Klatt}}, \bibinfo {author} {\bibfnamefont {B.}~\bibnamefont {Butz}},
  \bibinfo {author} {\bibfnamefont {M.~G.}\ \bibnamefont {Fischer}}, \bibinfo
  {author} {\bibfnamefont {S.~T.}\ \bibnamefont {Kelly}}, \bibinfo {author}
  {\bibfnamefont {E.}~\bibnamefont {Spiecker}}, \bibinfo {author}
  {\bibfnamefont {U.}~\bibnamefont {Steiner}},\ and\ \bibinfo {author}
  {\bibfnamefont {G.~E.}\ \bibnamefont {{Schr{\"o}der-Turk}}},\ }\href
  {https://doi.org/10.1126/sciadv.1603119} {\bibfield  {journal} {\bibinfo
  {journal} {Sci. Adv.}\ }\textbf {\bibinfo {volume} {3}},\ \bibinfo {pages}
  {e1603119} (\bibinfo {year} {2017})}\BibitemShut {NoStop}%
\bibitem [{\citenamefont {Sunada}(2008)}]{sunada_crystals_2008}%
  \BibitemOpen
  \bibfield  {author} {\bibinfo {author} {\bibfnamefont {T.}~\bibnamefont
  {Sunada}},\ }\href@noop {} {\bibfield  {journal} {\bibinfo  {journal} {Not.
  Am. Math. Soc.}\ }\textbf {\bibinfo {volume} {55}},\ \bibinfo {pages} {8}
  (\bibinfo {year} {2008})}\BibitemShut {NoStop}%
\bibitem [{\citenamefont {S{\"o}z{\"u}er}\ and\ \citenamefont
  {Haus}(1993)}]{sozuer_photonic_1993}%
  \BibitemOpen
  \bibfield  {author} {\bibinfo {author} {\bibfnamefont {H.~S.}\ \bibnamefont
  {S{\"o}z{\"u}er}}\ and\ \bibinfo {author} {\bibfnamefont {J.~W.}\
  \bibnamefont {Haus}},\ }\href {https://doi.org/10.1364/JOSAB.10.000296}
  {\bibfield  {journal} {\bibinfo  {journal} {J. Opt. Soc. Am. B}\ }\textbf
  {\bibinfo {volume} {10}},\ \bibinfo {pages} {296} (\bibinfo {year}
  {1993})}\BibitemShut {NoStop}%
\bibitem [{\citenamefont {Maldovan}\ and\ \citenamefont
  {Thomas}(2005)}]{maldovan_photonic_2005}%
  \BibitemOpen
  \bibfield  {author} {\bibinfo {author} {\bibfnamefont {M.}~\bibnamefont
  {Maldovan}}\ and\ \bibinfo {author} {\bibfnamefont {E.~L.}\ \bibnamefont
  {Thomas}},\ }\href {https://doi.org/10.1364/JOSAB.22.000466} {\bibfield
  {journal} {\bibinfo  {journal} {J. Opt. Soc. Am. B}\ }\textbf {\bibinfo
  {volume} {22}},\ \bibinfo {pages} {466} (\bibinfo {year} {2005})}\BibitemShut
  {NoStop}%
\bibitem [{\citenamefont {Barkema}\ and\ \citenamefont
  {Mousseau}(2000)}]{barkema_high-quality_2000}%
  \BibitemOpen
  \bibfield  {author} {\bibinfo {author} {\bibfnamefont {G.~T.}\ \bibnamefont
  {Barkema}}\ and\ \bibinfo {author} {\bibfnamefont {N.}~\bibnamefont
  {Mousseau}},\ }\href {https://doi.org/10.1103/PhysRevB.62.4985} {\bibfield
  {journal} {\bibinfo  {journal} {Phys. Rev. B}\ }\textbf {\bibinfo {volume}
  {62}},\ \bibinfo {pages} {4985} (\bibinfo {year} {2000})}\BibitemShut
  {NoStop}%
\bibitem [{\citenamefont {Torquato}\ and\ \citenamefont
  {Stillinger}(2003)}]{torquato_local_2003}%
  \BibitemOpen
  \bibfield  {author} {\bibinfo {author} {\bibfnamefont {S.}~\bibnamefont
  {Torquato}}\ and\ \bibinfo {author} {\bibfnamefont {F.~H.}\ \bibnamefont
  {Stillinger}},\ }\href {https://doi.org/10.1103/PhysRevE.68.041113}
  {\bibfield  {journal} {\bibinfo  {journal} {Phys. Rev. E}\ }\textbf {\bibinfo
  {volume} {68}},\ \bibinfo {pages} {041113} (\bibinfo {year}
  {2003})}\BibitemShut {NoStop}%
\bibitem [{\citenamefont {Torquato}(2010)}]{torquato_reformulation_2010}%
  \BibitemOpen
  \bibfield  {author} {\bibinfo {author} {\bibfnamefont {S.}~\bibnamefont
  {Torquato}},\ }\href {https://doi.org/10.1103/PhysRevE.82.056109} {\bibfield
  {journal} {\bibinfo  {journal} {Phys. Rev. E}\ }\textbf {\bibinfo {volume}
  {82}},\ \bibinfo {pages} {056109} (\bibinfo {year} {2010})}\BibitemShut
  {NoStop}%
\bibitem [{\citenamefont {Lloyd}(1982)}]{lloyd_least_1982}%
  \BibitemOpen
  \bibfield  {author} {\bibinfo {author} {\bibfnamefont {S.}~\bibnamefont
  {Lloyd}},\ }\href {https://doi.org/10.1109/TIT.1982.1056489} {\bibfield
  {journal} {\bibinfo  {journal} {IEEE Trans. Inf. Theory}\ }\textbf {\bibinfo
  {volume} {28}},\ \bibinfo {pages} {129} (\bibinfo {year} {1982})}\BibitemShut
  {NoStop}%
\bibitem [{\citenamefont {Liu}\ \emph {et~al.}(2009)\citenamefont {Liu},
  \citenamefont {Wang}, \citenamefont {L{\'e}vy}, \citenamefont {Sun},
  \citenamefont {Yan}, \citenamefont {Lu},\ and\ \citenamefont
  {Yang}}]{liu_centroidal_2009}%
  \BibitemOpen
  \bibfield  {author} {\bibinfo {author} {\bibfnamefont {Y.}~\bibnamefont
  {Liu}}, \bibinfo {author} {\bibfnamefont {W.}~\bibnamefont {Wang}}, \bibinfo
  {author} {\bibfnamefont {B.}~\bibnamefont {L{\'e}vy}}, \bibinfo {author}
  {\bibfnamefont {F.}~\bibnamefont {Sun}}, \bibinfo {author} {\bibfnamefont
  {D.-M.}\ \bibnamefont {Yan}}, \bibinfo {author} {\bibfnamefont
  {L.}~\bibnamefont {Lu}},\ and\ \bibinfo {author} {\bibfnamefont
  {C.}~\bibnamefont {Yang}},\ }\href {https://doi.org/10.1145/1559755.1559758}
  {\bibfield  {journal} {\bibinfo  {journal} {ACM Trans. Graph.}\ }\textbf
  {\bibinfo {volume} {28}},\ \bibinfo {pages} {101:1} (\bibinfo {year}
  {2009})}\BibitemShut {NoStop}%
\bibitem [{\citenamefont {Du}\ \emph {et~al.}(2010)\citenamefont {Du},
  \citenamefont {Gunzburger},\ and\ \citenamefont {Ju}}]{du_advances_2010}%
  \BibitemOpen
  \bibfield  {author} {\bibinfo {author} {\bibfnamefont {Q.}~\bibnamefont
  {Du}}, \bibinfo {author} {\bibfnamefont {M.}~\bibnamefont {Gunzburger}},\
  and\ \bibinfo {author} {\bibfnamefont {L.}~\bibnamefont {Ju}},\ }\href
  {https://doi.org/10.4208/nmtma.2010.32s.1} {\bibfield  {journal} {\bibinfo
  {journal} {Numer. Math.}\ }\textbf {\bibinfo {volume} {3}},\ \bibinfo {pages}
  {119} (\bibinfo {year} {2010})}\BibitemShut {NoStop}%
\bibitem [{\citenamefont {Zhang}\ \emph {et~al.}(2012)\citenamefont {Zhang},
  \citenamefont {Emelianenko},\ and\ \citenamefont {Du}}]{zhang_periodic_2012}%
  \BibitemOpen
  \bibfield  {author} {\bibinfo {author} {\bibfnamefont {J.}~\bibnamefont
  {Zhang}}, \bibinfo {author} {\bibfnamefont {M.}~\bibnamefont {Emelianenko}},\
  and\ \bibinfo {author} {\bibfnamefont {Q.}~\bibnamefont {Du}},\ }\href@noop
  {} {\bibfield  {journal} {\bibinfo  {journal} {Int. J. Numer. Anal. Model.}\
  }\textbf {\bibinfo {volume} {9}},\ \bibinfo {pages} {950} (\bibinfo {year}
  {2012})}\BibitemShut {NoStop}%
\bibitem [{\citenamefont {Ruscher}\ \emph {et~al.}(2015)\citenamefont
  {Ruscher}, \citenamefont {Baschnagel},\ and\ \citenamefont
  {Farago}}]{ruscher_voronoi_2015}%
  \BibitemOpen
  \bibfield  {author} {\bibinfo {author} {\bibfnamefont {C.}~\bibnamefont
  {Ruscher}}, \bibinfo {author} {\bibfnamefont {J.}~\bibnamefont
  {Baschnagel}},\ and\ \bibinfo {author} {\bibfnamefont {J.}~\bibnamefont
  {Farago}},\ }\href {https://doi.org/10.1209/0295-5075/112/66003} {\bibfield
  {journal} {\bibinfo  {journal} {Europhys. Lett.}\ }\textbf {\bibinfo {volume}
  {112}},\ \bibinfo {pages} {66003} (\bibinfo {year} {2015})}\BibitemShut
  {NoStop}%
\bibitem [{\citenamefont {Ruscher}\ \emph {et~al.}(2021)\citenamefont
  {Ruscher}, \citenamefont {Ciarella}, \citenamefont {Luo}, \citenamefont
  {Janssen}, \citenamefont {Farago},\ and\ \citenamefont
  {Baschnagel}}]{ruscher_glassy_2020}%
  \BibitemOpen
  \bibfield  {author} {\bibinfo {author} {\bibfnamefont {C.}~\bibnamefont
  {Ruscher}}, \bibinfo {author} {\bibfnamefont {S.}~\bibnamefont {Ciarella}},
  \bibinfo {author} {\bibfnamefont {C.}~\bibnamefont {Luo}}, \bibinfo {author}
  {\bibfnamefont {L.~M.~C.}\ \bibnamefont {Janssen}}, \bibinfo {author}
  {\bibfnamefont {J.}~\bibnamefont {Farago}},\ and\ \bibinfo {author}
  {\bibfnamefont {J.}~\bibnamefont {Baschnagel}},\ }\href
  {https://doi.org/10.1088/1361-648X/abc4cc} {\bibfield  {journal} {\bibinfo
  {journal} {J. Phys. Condens. Matter}\ }\textbf {\bibinfo {volume} {33}},\
  \bibinfo {pages} {064001} (\bibinfo {year} {2021})}\BibitemShut {NoStop}%
\bibitem [{\citenamefont {Hain}\ \emph {et~al.}(2020)\citenamefont {Hain},
  \citenamefont {Klatt},\ and\ \citenamefont
  {{Schr{\"o}der-Turk}}}]{hain_low-temperature_2020}%
  \BibitemOpen
  \bibfield  {author} {\bibinfo {author} {\bibfnamefont {T.~M.}\ \bibnamefont
  {Hain}}, \bibinfo {author} {\bibfnamefont {M.~A.}\ \bibnamefont {Klatt}},\
  and\ \bibinfo {author} {\bibfnamefont {G.~E.}\ \bibnamefont
  {{Schr{\"o}der-Turk}}},\ }\href {https://doi.org/10.1063/5.0029301}
  {\bibfield  {journal} {\bibinfo  {journal} {J. Chem. Phys.}\ }\textbf
  {\bibinfo {volume} {153}},\ \bibinfo {pages} {234505} (\bibinfo {year}
  {2020})}\BibitemShut {NoStop}%
\bibitem [{\citenamefont {Haberko}\ and\ \citenamefont
  {Scheffold}(2013)}]{haberko_fabrication_2013}%
  \BibitemOpen
  \bibfield  {author} {\bibinfo {author} {\bibfnamefont {J.}~\bibnamefont
  {Haberko}}\ and\ \bibinfo {author} {\bibfnamefont {F.}~\bibnamefont
  {Scheffold}},\ }\href {https://doi.org/10.1364/OE.21.001057} {\bibfield
  {journal} {\bibinfo  {journal} {Opt. Express}\ }\textbf {\bibinfo {volume}
  {21}},\ \bibinfo {pages} {1057} (\bibinfo {year} {2013})}\BibitemShut
  {NoStop}%
\end{thebibliography}%


\begin{thebibliography}{13}%
\makeatletter
\providecommand \@ifxundefined [1]{%
 \@ifx{#1\undefined}
}%
\providecommand \@ifnum [1]{%
 \ifnum #1\expandafter \@firstoftwo
 \else \expandafter \@secondoftwo
 \fi
}%
\providecommand \@ifx [1]{%
 \ifx #1\expandafter \@firstoftwo
 \else \expandafter \@secondoftwo
 \fi
}%
\providecommand \natexlab [1]{#1}%
\providecommand \enquote  [1]{``#1''}%
\providecommand \bibnamefont  [1]{#1}%
\providecommand \bibfnamefont [1]{#1}%
\providecommand \citenamefont [1]{#1}%
\providecommand \href@noop [0]{\@secondoftwo}%
\providecommand \href [0]{\begingroup \@sanitize@url \@href}%
\providecommand \@href[1]{\@@startlink{#1}\@@href}%
\providecommand \@@href[1]{\endgroup#1\@@endlink}%
\providecommand \@sanitize@url [0]{\catcode `\\12\catcode `\$12\catcode
  `\&12\catcode `\#12\catcode `\^12\catcode `\_12\catcode `\%12\relax}%
\providecommand \@@startlink[1]{}%
\providecommand \@@endlink[0]{}%
\providecommand \url  [0]{\begingroup\@sanitize@url \@url }%
\providecommand \@url [1]{\endgroup\@href {#1}{\urlprefix }}%
\providecommand \urlprefix  [0]{URL }%
\providecommand \Eprint [0]{\href }%
\providecommand \doibase [0]{https://doi.org/}%
\providecommand \selectlanguage [0]{\@gobble}%
\providecommand \bibinfo  [0]{\@secondoftwo}%
\providecommand \bibfield  [0]{\@secondoftwo}%
\providecommand \translation [1]{[#1]}%
\providecommand \BibitemOpen [0]{}%
\providecommand \bibitemStop [0]{}%
\providecommand \bibitemNoStop [0]{.\EOS\space}%
\providecommand \EOS [0]{\spacefactor3000\relax}%
\providecommand \BibitemShut  [1]{\csname bibitem#1\endcsname}%
\let\auto@bib@innerbib\@empty
%</preamble>
\bibitem [{\citenamefont {Hejna}\ \emph {et~al.}(2013)\citenamefont {Hejna},
  \citenamefont {Steinhardt},\ and\ \citenamefont
  {Torquato}}]{hejna_nearly_2013}%
  \BibitemOpen
  \bibfield  {author} {\bibinfo {author} {\bibfnamefont {M.}~\bibnamefont
  {Hejna}}, \bibinfo {author} {\bibfnamefont {P.~J.}\ \bibnamefont
  {Steinhardt}},\ and\ \bibinfo {author} {\bibfnamefont {S.}~\bibnamefont
  {Torquato}},\ }\href {https://doi.org/10.1103/PhysRevB.87.245204} {\bibfield
  {journal} {\bibinfo  {journal} {Phys. Rev. B}\ }\textbf {\bibinfo {volume}
  {87}},\ \bibinfo {pages} {245204} (\bibinfo {year} {2013})}\BibitemShut
  {NoStop}%
\bibitem [{\citenamefont {Deringer}\ \emph {et~al.}(2018)\citenamefont
  {Deringer}, \citenamefont {Bernstein}, \citenamefont {Bart{\'o}k},
  \citenamefont {Cliffe}, \citenamefont {Kerber}, \citenamefont {Marbella},
  \citenamefont {Grey}, \citenamefont {Elliott},\ and\ \citenamefont
  {Cs{\'a}nyi}}]{deringer_realistic_2018}%
  \BibitemOpen
  \bibfield  {author} {\bibinfo {author} {\bibfnamefont {V.~L.}\ \bibnamefont
  {Deringer}}, \bibinfo {author} {\bibfnamefont {N.}~\bibnamefont {Bernstein}},
  \bibinfo {author} {\bibfnamefont {A.~P.}\ \bibnamefont {Bart{\'o}k}},
  \bibinfo {author} {\bibfnamefont {M.~J.}\ \bibnamefont {Cliffe}}, \bibinfo
  {author} {\bibfnamefont {R.~N.}\ \bibnamefont {Kerber}}, \bibinfo {author}
  {\bibfnamefont {L.~E.}\ \bibnamefont {Marbella}}, \bibinfo {author}
  {\bibfnamefont {C.~P.}\ \bibnamefont {Grey}}, \bibinfo {author}
  {\bibfnamefont {S.~R.}\ \bibnamefont {Elliott}},\ and\ \bibinfo {author}
  {\bibfnamefont {G.}~\bibnamefont {Cs{\'a}nyi}},\ }\href
  {https://doi.org/10.1021/acs.jpclett.8b00902} {\bibfield  {journal} {\bibinfo
   {journal} {J. Phys. Chem. Lett.}\ }\textbf {\bibinfo {volume} {9}},\
  \bibinfo {pages} {2879} (\bibinfo {year} {2018})}\BibitemShut {NoStop}%
\bibitem [{\citenamefont {Klatt}\ \emph {et~al.}(2019)\citenamefont {Klatt},
  \citenamefont {Lovri{\'c}}, \citenamefont {Chen}, \citenamefont {Kapfer},
  \citenamefont {Schaller}, \citenamefont {Sch{\"o}nh{\"o}fer}, \citenamefont
  {Gardiner}, \citenamefont {Smith}, \citenamefont {{Schr{\"o}der-Turk}},\ and\
  \citenamefont {Torquato}}]{klatt_universal_2019}%
  \BibitemOpen
  \bibfield  {author} {\bibinfo {author} {\bibfnamefont {M.~A.}\ \bibnamefont
  {Klatt}}, \bibinfo {author} {\bibfnamefont {J.}~\bibnamefont {Lovri{\'c}}},
  \bibinfo {author} {\bibfnamefont {D.}~\bibnamefont {Chen}}, \bibinfo {author}
  {\bibfnamefont {S.~C.}\ \bibnamefont {Kapfer}}, \bibinfo {author}
  {\bibfnamefont {F.~M.}\ \bibnamefont {Schaller}}, \bibinfo {author}
  {\bibfnamefont {P.~W.~A.}\ \bibnamefont {Sch{\"o}nh{\"o}fer}}, \bibinfo
  {author} {\bibfnamefont {B.~S.}\ \bibnamefont {Gardiner}}, \bibinfo {author}
  {\bibfnamefont {A.-S.}\ \bibnamefont {Smith}}, \bibinfo {author}
  {\bibfnamefont {G.~E.}\ \bibnamefont {{Schr{\"o}der-Turk}}},\ and\ \bibinfo
  {author} {\bibfnamefont {S.}~\bibnamefont {Torquato}},\ }\href
  {https://doi.org/10.1038/s41467-019-08360-5} {\bibfield  {journal} {\bibinfo
  {journal} {Nat. Commun.}\ }\textbf {\bibinfo {volume} {10}},\ \bibinfo
  {pages} {811} (\bibinfo {year} {2019})}\BibitemShut {NoStop}%
\bibitem [{\citenamefont {Liu}\ \emph {et~al.}(2009)\citenamefont {Liu},
  \citenamefont {Wang}, \citenamefont {L{\'e}vy}, \citenamefont {Sun},
  \citenamefont {Yan}, \citenamefont {Lu},\ and\ \citenamefont
  {Yang}}]{liu_centroidal_2009}%
  \BibitemOpen
  \bibfield  {author} {\bibinfo {author} {\bibfnamefont {Y.}~\bibnamefont
  {Liu}}, \bibinfo {author} {\bibfnamefont {W.}~\bibnamefont {Wang}}, \bibinfo
  {author} {\bibfnamefont {B.}~\bibnamefont {L{\'e}vy}}, \bibinfo {author}
  {\bibfnamefont {F.}~\bibnamefont {Sun}}, \bibinfo {author} {\bibfnamefont
  {D.-M.}\ \bibnamefont {Yan}}, \bibinfo {author} {\bibfnamefont
  {L.}~\bibnamefont {Lu}},\ and\ \bibinfo {author} {\bibfnamefont
  {C.}~\bibnamefont {Yang}},\ }\href {https://doi.org/10.1145/1559755.1559758}
  {\bibfield  {journal} {\bibinfo  {journal} {ACM Trans. Graph.}\ }\textbf
  {\bibinfo {volume} {28}},\ \bibinfo {pages} {101:1} (\bibinfo {year}
  {2009})}\BibitemShut {NoStop}%
\bibitem [{\citenamefont {Du}\ \emph {et~al.}(2010)\citenamefont {Du},
  \citenamefont {Gunzburger},\ and\ \citenamefont {Ju}}]{du_advances_2010}%
  \BibitemOpen
  \bibfield  {author} {\bibinfo {author} {\bibfnamefont {Q.}~\bibnamefont
  {Du}}, \bibinfo {author} {\bibfnamefont {M.}~\bibnamefont {Gunzburger}},\
  and\ \bibinfo {author} {\bibfnamefont {L.}~\bibnamefont {Ju}},\ }\href
  {https://doi.org/10.4208/nmtma.2010.32s.1} {\bibfield  {journal} {\bibinfo
  {journal} {Numer. Math.}\ }\textbf {\bibinfo {volume} {3}},\ \bibinfo {pages}
  {119} (\bibinfo {year} {2010})}\BibitemShut {NoStop}%
\bibitem [{\citenamefont {Torquato}(2010)}]{torquato_reformulation_2010}%
  \BibitemOpen
  \bibfield  {author} {\bibinfo {author} {\bibfnamefont {S.}~\bibnamefont
  {Torquato}},\ }\href {https://doi.org/10.1103/PhysRevE.82.056109} {\bibfield
  {journal} {\bibinfo  {journal} {Phys. Rev. E}\ }\textbf {\bibinfo {volume}
  {82}},\ \bibinfo {pages} {056109} (\bibinfo {year} {2010})}\BibitemShut
  {NoStop}%
\bibitem [{\citenamefont {Zhang}\ \emph {et~al.}(2012)\citenamefont {Zhang},
  \citenamefont {Emelianenko},\ and\ \citenamefont {Du}}]{zhang_periodic_2012}%
  \BibitemOpen
  \bibfield  {author} {\bibinfo {author} {\bibfnamefont {J.}~\bibnamefont
  {Zhang}}, \bibinfo {author} {\bibfnamefont {M.}~\bibnamefont {Emelianenko}},\
  and\ \bibinfo {author} {\bibfnamefont {Q.}~\bibnamefont {Du}},\ }\href@noop
  {} {\bibfield  {journal} {\bibinfo  {journal} {Int. J. Numer. Anal. Model.}\
  }\textbf {\bibinfo {volume} {9}},\ \bibinfo {pages} {950} (\bibinfo {year}
  {2012})}\BibitemShut {NoStop}%
\bibitem [{\citenamefont {Ruscher}\ \emph {et~al.}(2015)\citenamefont
  {Ruscher}, \citenamefont {Baschnagel},\ and\ \citenamefont
  {Farago}}]{ruscher_voronoi_2015}%
  \BibitemOpen
  \bibfield  {author} {\bibinfo {author} {\bibfnamefont {C.}~\bibnamefont
  {Ruscher}}, \bibinfo {author} {\bibfnamefont {J.}~\bibnamefont
  {Baschnagel}},\ and\ \bibinfo {author} {\bibfnamefont {J.}~\bibnamefont
  {Farago}},\ }\href {https://doi.org/10.1209/0295-5075/112/66003} {\bibfield
  {journal} {\bibinfo  {journal} {Europhys. Lett.}\ }\textbf {\bibinfo {volume}
  {112}},\ \bibinfo {pages} {66003} (\bibinfo {year} {2015})}\BibitemShut
  {NoStop}%
\bibitem [{\citenamefont {Ruscher}\ \emph {et~al.}(2021)\citenamefont
  {Ruscher}, \citenamefont {Ciarella}, \citenamefont {Luo}, \citenamefont
  {Janssen}, \citenamefont {Farago},\ and\ \citenamefont
  {Baschnagel}}]{ruscher_glassy_2020}%
  \BibitemOpen
  \bibfield  {author} {\bibinfo {author} {\bibfnamefont {C.}~\bibnamefont
  {Ruscher}}, \bibinfo {author} {\bibfnamefont {S.}~\bibnamefont {Ciarella}},
  \bibinfo {author} {\bibfnamefont {C.}~\bibnamefont {Luo}}, \bibinfo {author}
  {\bibfnamefont {L.~M.~C.}\ \bibnamefont {Janssen}}, \bibinfo {author}
  {\bibfnamefont {J.}~\bibnamefont {Farago}},\ and\ \bibinfo {author}
  {\bibfnamefont {J.}~\bibnamefont {Baschnagel}},\ }\href
  {https://doi.org/10.1088/1361-648X/abc4cc} {\bibfield  {journal} {\bibinfo
  {journal} {J. Phys. Condens. Matter}\ }\textbf {\bibinfo {volume} {33}},\
  \bibinfo {pages} {064001} (\bibinfo {year} {2021})}\BibitemShut {NoStop}%
\bibitem [{\citenamefont {Hain}\ \emph {et~al.}(2020)\citenamefont {Hain},
  \citenamefont {Klatt},\ and\ \citenamefont
  {{Schr{\"o}der-Turk}}}]{hain_low-temperature_2020}%
  \BibitemOpen
  \bibfield  {author} {\bibinfo {author} {\bibfnamefont {T.~M.}\ \bibnamefont
  {Hain}}, \bibinfo {author} {\bibfnamefont {M.~A.}\ \bibnamefont {Klatt}},\
  and\ \bibinfo {author} {\bibfnamefont {G.~E.}\ \bibnamefont
  {{Schr{\"o}der-Turk}}},\ }\href {https://doi.org/10.1063/5.0029301}
  {\bibfield  {journal} {\bibinfo  {journal} {J. Chem. Phys.}\ }\textbf
  {\bibinfo {volume} {153}},\ \bibinfo {pages} {234505} (\bibinfo {year}
  {2020})}\BibitemShut {NoStop}%
\bibitem [{\citenamefont {Lloyd}(1982)}]{lloyd_least_1982}%
  \BibitemOpen
  \bibfield  {author} {\bibinfo {author} {\bibfnamefont {S.}~\bibnamefont
  {Lloyd}},\ }\href {https://doi.org/10.1109/TIT.1982.1056489} {\bibfield
  {journal} {\bibinfo  {journal} {IEEE Trans. Inf. Theory}\ }\textbf {\bibinfo
  {volume} {28}},\ \bibinfo {pages} {129} (\bibinfo {year} {1982})}\BibitemShut
  {NoStop}%
\bibitem [{\citenamefont {Barkema}\ and\ \citenamefont
  {Mousseau}(2000)}]{barkema_high-quality_2000}%
  \BibitemOpen
  \bibfield  {author} {\bibinfo {author} {\bibfnamefont {G.~T.}\ \bibnamefont
  {Barkema}}\ and\ \bibinfo {author} {\bibfnamefont {N.}~\bibnamefont
  {Mousseau}},\ }\href {https://doi.org/10.1103/PhysRevB.62.4985} {\bibfield
  {journal} {\bibinfo  {journal} {Phys. Rev. B}\ }\textbf {\bibinfo {volume}
  {62}},\ \bibinfo {pages} {4985} (\bibinfo {year} {2000})}\BibitemShut
  {NoStop}%
\bibitem [{\citenamefont {Johnson}\ and\ \citenamefont
  {Joannopoulos}(2001)}]{johnson_block-iterative_2001}%
  \BibitemOpen
  \bibfield  {author} {\bibinfo {author} {\bibfnamefont {S.}~\bibnamefont
  {Johnson}}\ and\ \bibinfo {author} {\bibfnamefont {J.}~\bibnamefont
  {Joannopoulos}},\ }\href {https://doi.org/10.1364/OE.8.000173} {\bibfield
  {journal} {\bibinfo  {journal} {Opt. Express}\ }\textbf {\bibinfo {volume}
  {8}},\ \bibinfo {pages} {173} (\bibinfo {year} {2001})}\BibitemShut {NoStop}%
\end{thebibliography}%

\end{document}

\title{Supplemental Material:\texorpdfstring{\\}{}
Gap Sensitivity Reveals Universal Behaviors in
Optimized Photonic \texorpdfstring{\\}{} Crystal and Disordered Networks
}

\author{Michael A. Klatt}
\affiliation{Department of Physics, Princeton University, Princeton, New Jersey 08544, USA}
\affiliation{Institut für Theoretische Physik, University of Erlangen-Nürnberg, Staudtstr. 7, 91058 Erlangen, Germany}
\author{Paul J. Steinhardt} 
\affiliation{Department of Physics, Princeton University, Princeton, NJ 08544, USA}
\author{Salvatore Torquato}
\affiliation{Department of Physics, Princeton University, Princeton, NJ 08544, USA}
\affiliation{Department of Chemistry, Princeton Institute for the Science and Technology of Materials, and Program in Applied and Computational Mathematics, Princeton University, Princeton, New Jersey 08544, USA}
\date{\today}
\maketitle

\subsection{S1.~Simulation details}

Here we provide additional parameters for our network models and the 
computation of their 
(optimal) photonic band gaps (PBGs).

\subsubsection{Network models}

For the nearly hyperuniform network (NHN), we used the most annealed 
sample with 1000 vertices from Ref.~\cite{hejna_nearly_2013}.
For the network based on the molecular dynamics (MD) quench, we used the 
sample~\verb+GAP-MD_quench_1e11_lastMDstep+ published together with 
Ref.~\cite{deringer_realistic_2018}.
The definition of our quantizer-based networks (QBNs) is based on the 
amorphous inherent structures of the quantizer energy from 
Ref.~\cite{klatt_universal_2019}; for references on the quantizer energy, see 
Refs.~\cite{liu_centroidal_2009,
du_advances_2010,
torquato_reformulation_2010,
zhang_periodic_2012,
ruscher_voronoi_2015,
ruscher_glassy_2020,
hain_low-temperature_2020,
klatt_universal_2019}.
To construct our QBN, we start from a binomial point process (that is, a 
snapshot of the ideal gas in the canonical ensemble) and minimize the 
quantizer energy using the so-called Lloyd 
algorithm~\cite{lloyd_least_1982}:
In each step, the Voronoi center of each cell is replaced by the center 
of mass of the cell.
The algorithm converges to an amorphous inherent 
structure~\cite{klatt_universal_2019}.
Here, we apply 10,000 steps of the Lloyd algorithm.
The network is then constructed using the Delaunay tessellation as 
explained in the main text.
Our definition of the QBNs allows the simulation of extended networks 
with a million vertices.
For computational reasons, we here simulated a sample with 444 
vertices.

The unit of length was chosen for each sample such that the number of 
vertices equals the volume of the sample.
This choice corresponds to a unit number density $\rho=1$ for the vertices.
The number of vertices per sample is 1000 for the NHN, 444 for the 
QBN, 512 for the MD quench, and 216 for the perturbed diamond networks.
To check for system size effects, we also analyzed a continuous random 
network (CRN) by Barkema and Mousseau~\cite{barkema_high-quality_2000} 
with 216 vertices and a perturbed diamond network ($a=0.2$) with 1000 
vertices; see Fig.~\ref{fig:sys-size}.

A single sample of a perturbed diamond or disordered network can have 
several complete PBGs between different bands.
Typically, one of these PBGs is much larger than the others; see video~S2
and Fig.~\ref{fig:bst}.
For each sample and each value of $\alpha$, we optimize the PBG for 
which we find the largest value of $\Delta(\alpha)$ (with few exceptions 
as explained below).
The number of bands below this largest PBG is 1000 for the NHN, 509 for the MD quench, 444 for 
the QBN for $\alpha > 9.3$ (for $\alpha \leq 9.3$, we used the same band number although the PBG at 
band number 442 is slightly larger), 1000 for the perturbed diamond network 
with 1000 vertices, 216 for the perturbed diamond networks with 216 
vertices and with $a=0.1$, 0.2, and 0.3, and 215 for the perturbed diamond 
network with $a=0.4$.
For the last sample, we found several small complete PBGs of roughly the 
same size so that the number of bands below the largest gap can differ 
for different values of $R$ and $\alpha$.
The same applies to our perturbed CRN with $a=0.3$.

\subsubsection{Parameters}
The resolution $\mathcal{R}$, which is defined in the MIT Photonic Bands
(\textsc{MPB}) software package as the number of voxels per unit
length, was set to the following values during the optimization of the
radii:
\begin{itemize}
  \item $\mathcal{R}=16$ for the disordered networks and perturbed diamond networks,
  \item $\mathcal{R}=20$ for the diamond, hexagonal-diamond, and Laves networks, and
  \item $\mathcal{R}=80$ for the simple cubic (SC) network.
\end{itemize}
For the final values of Figs.~2 and 3 in the main text, as well as for 
Fig.~\ref{fig:sys-size},
we used the following resolutions: 
\begin{itemize}
  \item $\mathcal{R}=16$ for the perturbed diamond network with 1000 vertices,
  \item $\mathcal{R}=20$ for the disordered networks and perturbed diamond networks,
  \item $\mathcal{R}=80$ for the diamond, hexagonal-diamond, and Laves networks, and
  \item $\mathcal{R}=160$ for the simple cubic (SC) network.
\end{itemize}
The number of $k$ points for which we computed the eigenfrequencies are
\begin{itemize}
  \item 10 for the NHN, MD quench, and the QBN,
  \item 22 for the perturbed diamond networks and the perturbed CRNs.
\end{itemize}
For the disordered networks, the $k$ points always include the high 
symmetry points (of the simple cubic simulation box).
For the crystal networks, we computed the eigenfrequencies at 8 points
between pairs of high symmetry points.
The tolerance of the \textsc{MPB} eigensolver was $10^{-5}$.

The calculations at strong contrasts are more prone to voxelization 
errors because of the smaller values of the optimal radii.
The estimation of the optimal radii is also more difficult at strong 
contrasts because the optimal gap size as a function of the radius $R$ 
becomes a skewed function (even close to the optimum).
Finally, single defects in disordered or perturbed networks can 
strongly influence the optimal gap size.
We, therefore, used the following improved parameters for some of the 
calculations at strong contrasts (to check their influence on the 
results, which was not found to be strong).
To determine the optimal radii for the crystal networks, we used a 
resolution of 80 for the diamond network with $\alpha \geq 20$ and the 
laves network with $\alpha\geq 18$. Moreover, we used a resolution of 20 
for the CRN with $\alpha=8.3$, 11.3, and 13.0.

Finally, in our study of system size effects in Fig.~\ref{fig:sys-size}, 
the computation for the perturbed diamond network with 1000 vertices was 
computationally particularly expensive because of the large number of 
$k$-points that have to be considered (because the perturbed diamond 
network has an anisotropic band structure---in contrast to the NHN).
Therefore, we did not optimize the radii separately but used the optimal 
radii of the perturbed diamond network with 216 vertices.
We checked for $\alpha = 4.8$ and 13 that these radii are too a good 
approximation also optimal for the network with 1000 vertices.
We also checked that the resolution $\mathcal{R}=16$ was sufficient for 
our calculations by comparing the results for two stop gaps with the 
corresponding results for a resolution $\mathcal{R}=20$.
The differences in the gap-to-midgap ratios were about 0.1\%, that is, 
within the accuracy of our study.

\subsubsection{Volume fractions}

We estimate the volume fractions $\phi$ of our networks using the 
dielectric filling fraction implemented in MPB.
The MPB software smooths the discontinuous dielectric function at the
resolution of the grid (to avoid convergence problems caused by the
discretization)~\cite{johnson_block-iterative_2001}.
The dielectric filling fraction is then defined as:
\begin{align}
  \varphi := \frac{\langle\varepsilon\rangle-\min\varepsilon}{\max\varepsilon-\min\varepsilon},
  \label{eq:filling}
\end{align}
where $\langle\varepsilon\rangle$, $\min\varepsilon$, and
$\max\varepsilon$ are the average, minimal, and maximal dielectric
constant over all voxels.
For a two-phase medium that consists of a high and a low dielectric
material, the filling fraction $\varphi$ is equivalent to the volume
fraction $\phi$ of the high dielectric phase.

\bibliography{gap-sensitivity}

% #################################################################### %

\newpage
\subsection{S2.~Photonic band structures}

\begin{figure}[h]
  \centering
  \hbox{}\hfill%
  \includegraphics[width=0.40\textwidth]{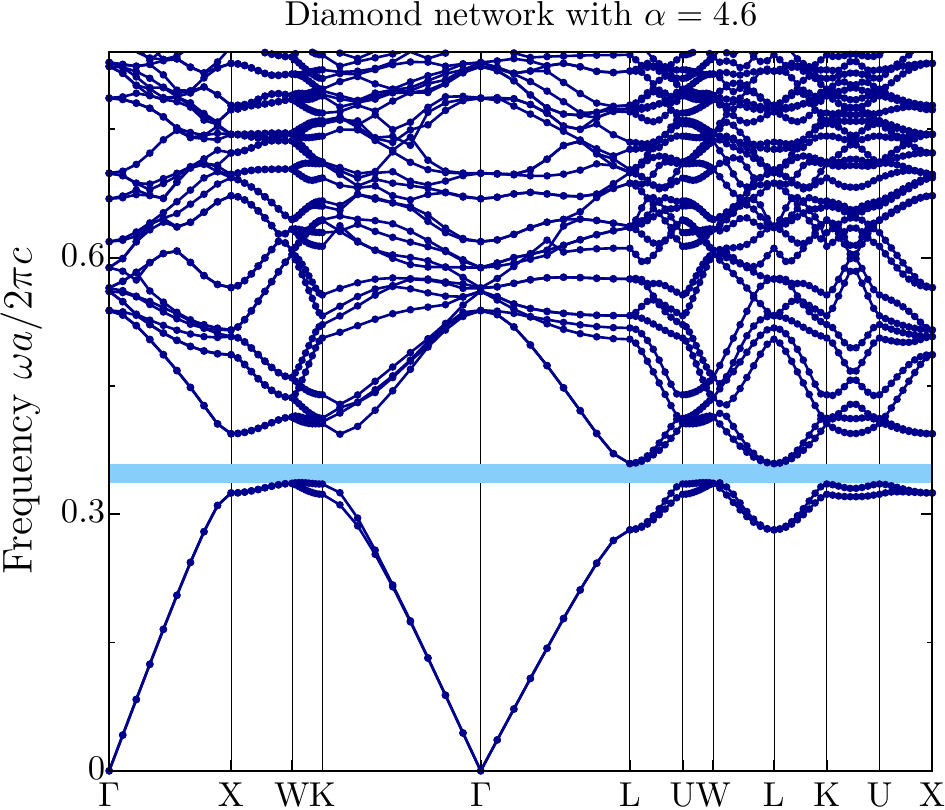}\hfill%
  \includegraphics[width=0.40\textwidth]{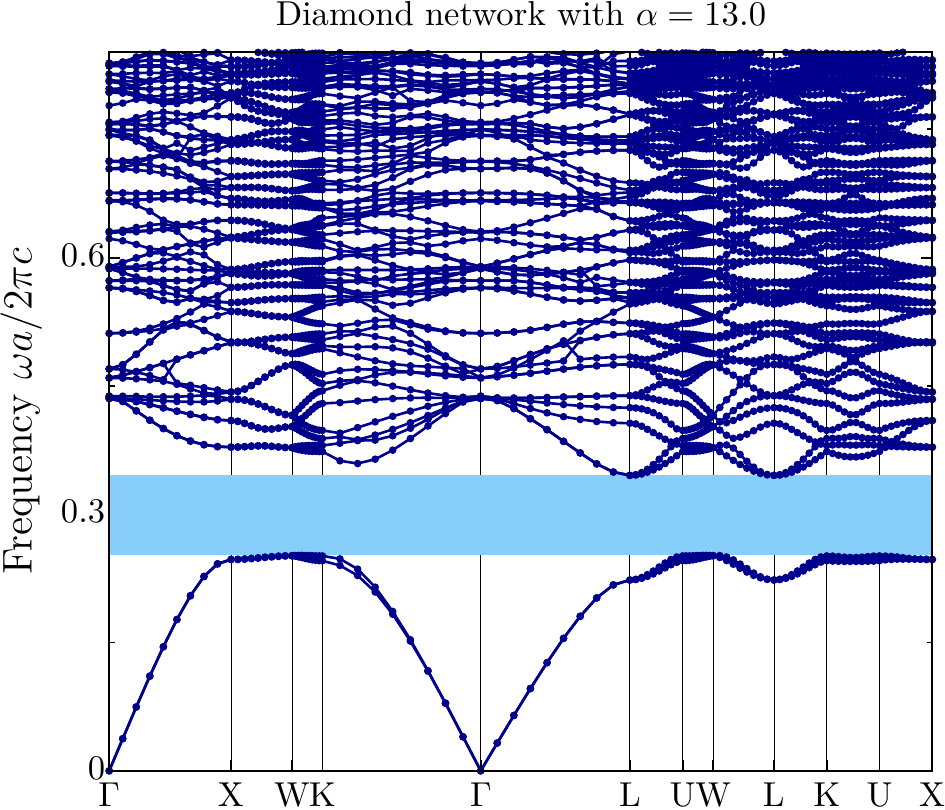}
  \hfill\hbox{}

  \hbox{}\hfill%
  \includegraphics[width=0.40\textwidth]{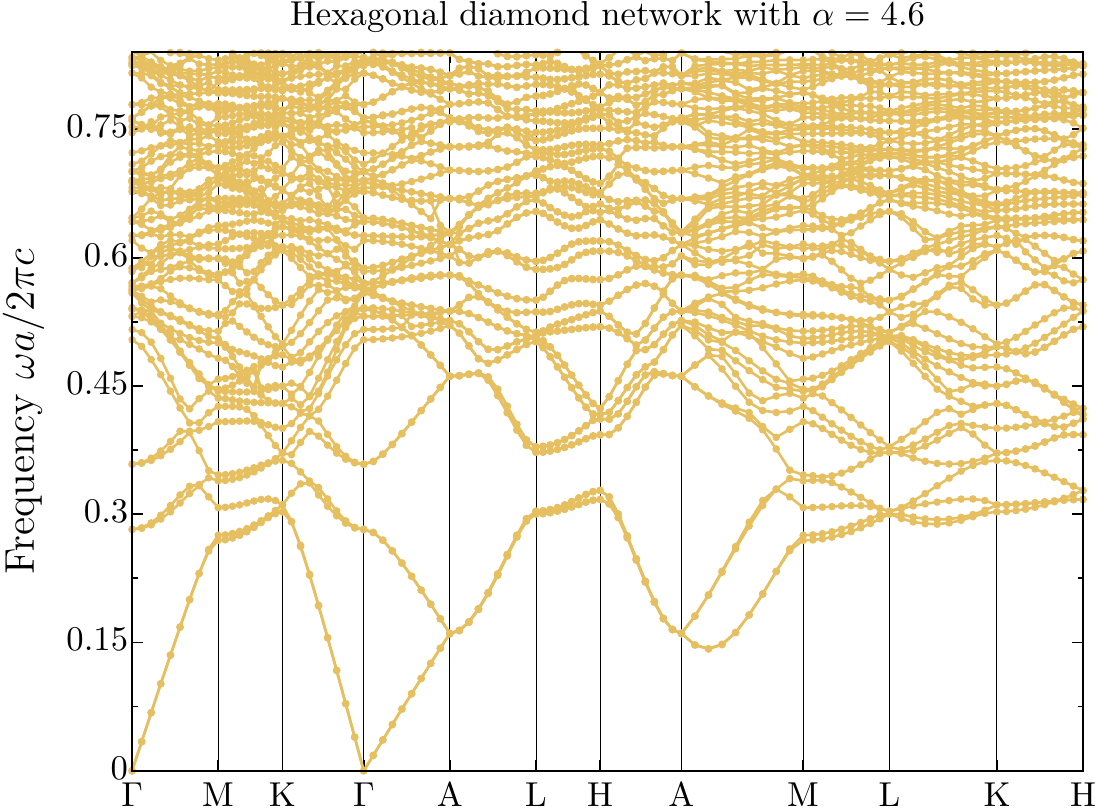}\hfill%
  \includegraphics[width=0.40\textwidth]{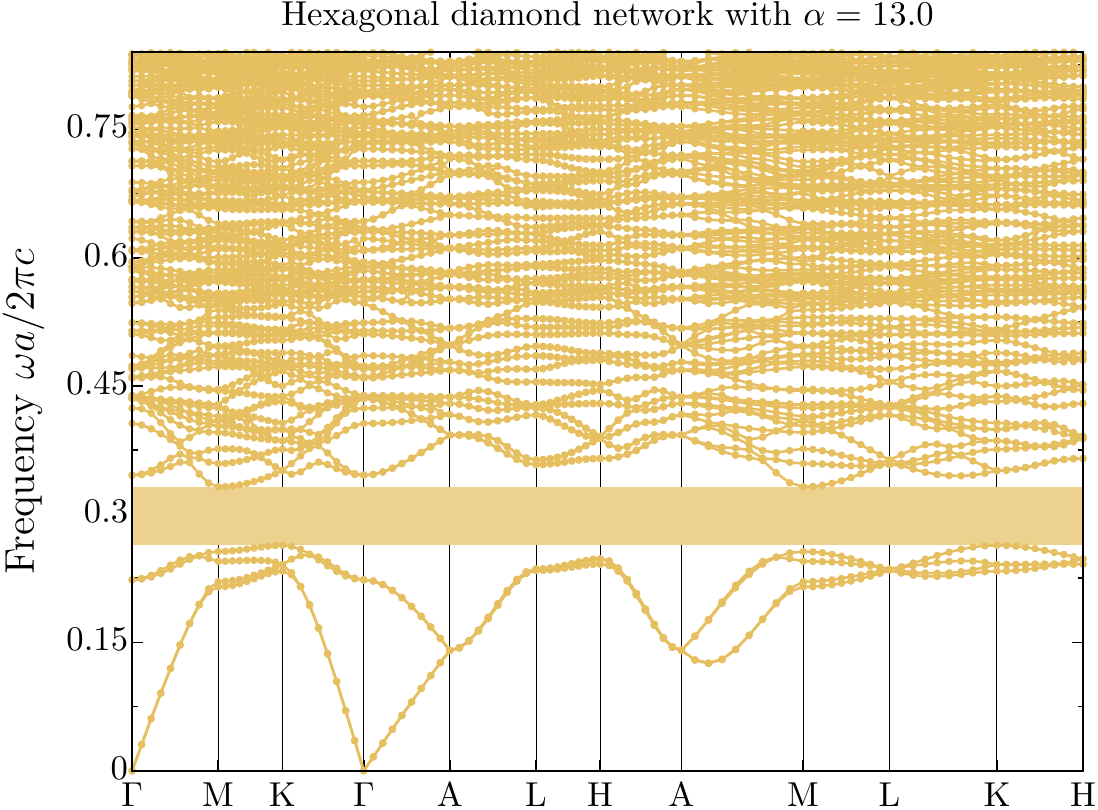}
  \hfill\hbox{}

  \hbox{}\hfill%
  \includegraphics[width=0.40\textwidth]{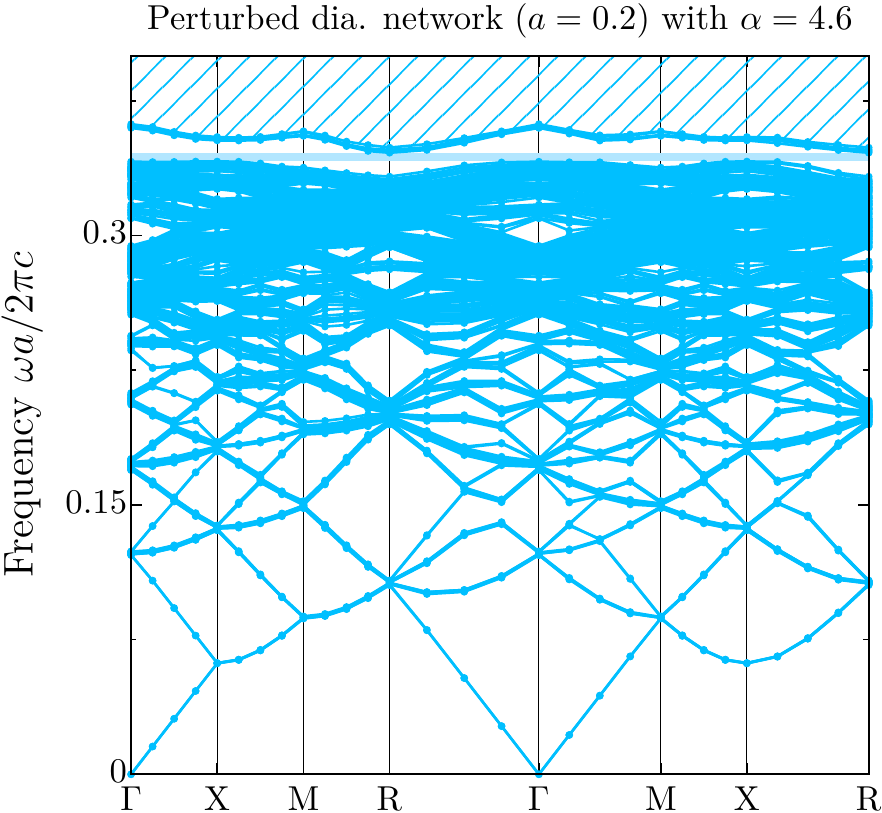}\hfill%
  \includegraphics[width=0.40\textwidth]{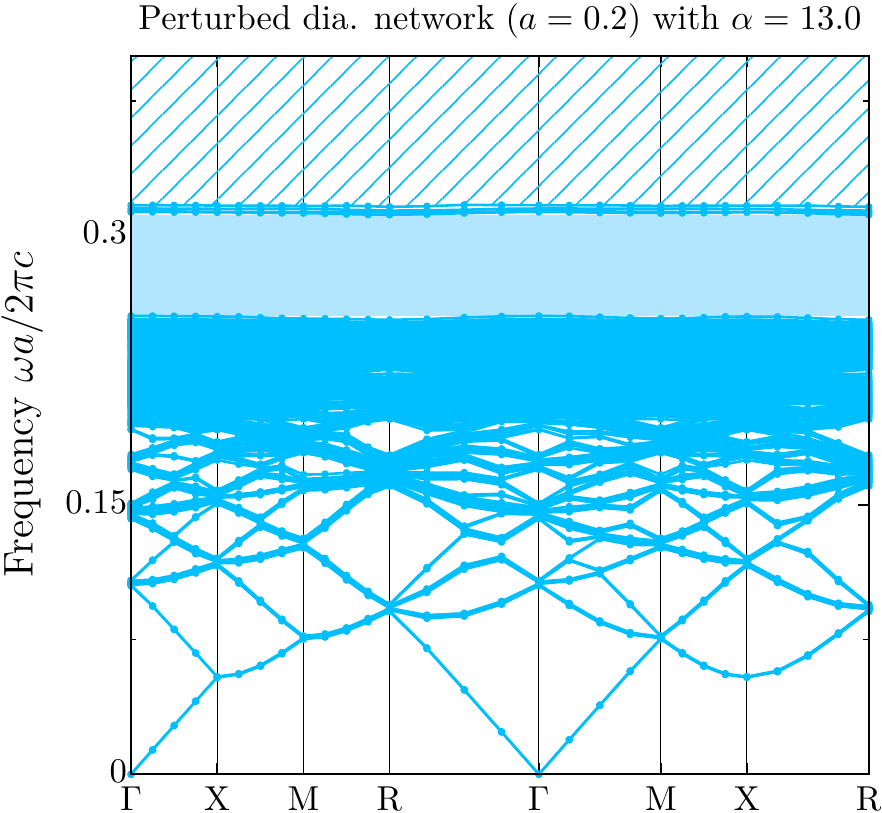}
  \hfill\hbox{}
  \caption{Photonic band structures for some of the (anisotropic) 
  crystal and perturbed crystal networks at different dielectric 
  contrasts $\alpha$; see also video~S2.}
  \label{fig:bst}
\end{figure}

\setcounter{figure}{0}

\begin{figure}[p]
%\ContinuedFloat
  \centering
  \hbox{}\hfill%
  \includegraphics[height=7.1cm]{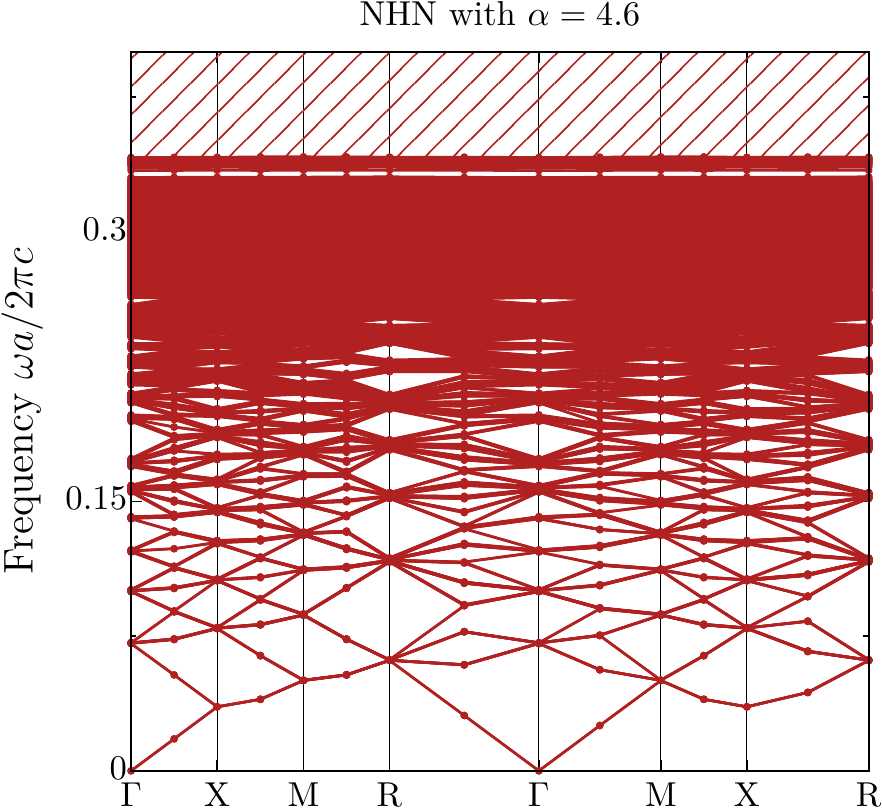}\hfill%
  \includegraphics[height=7.1cm]{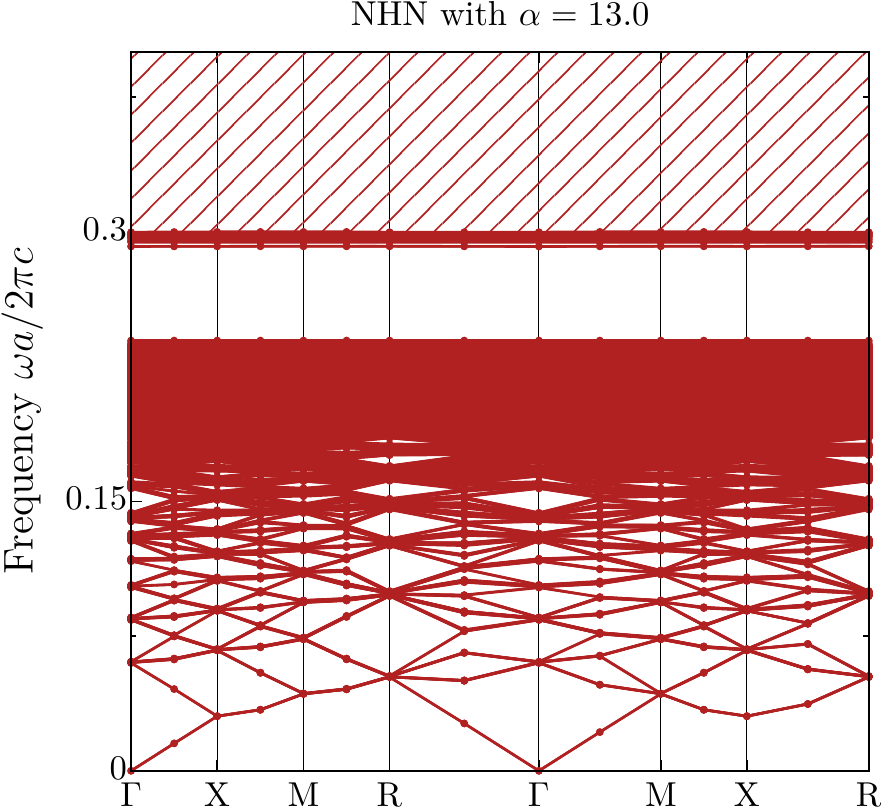}
  \hfill\hbox{}

  \hbox{}\hfill%
  \includegraphics[height=7.1cm]{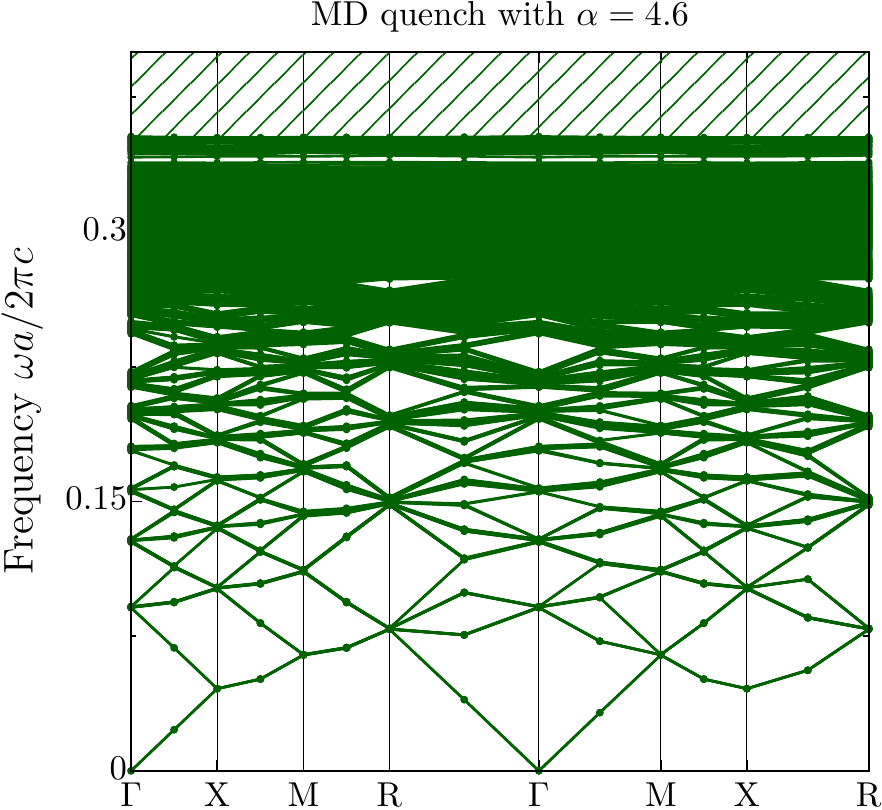}\hfill%
  \includegraphics[height=7.1cm]{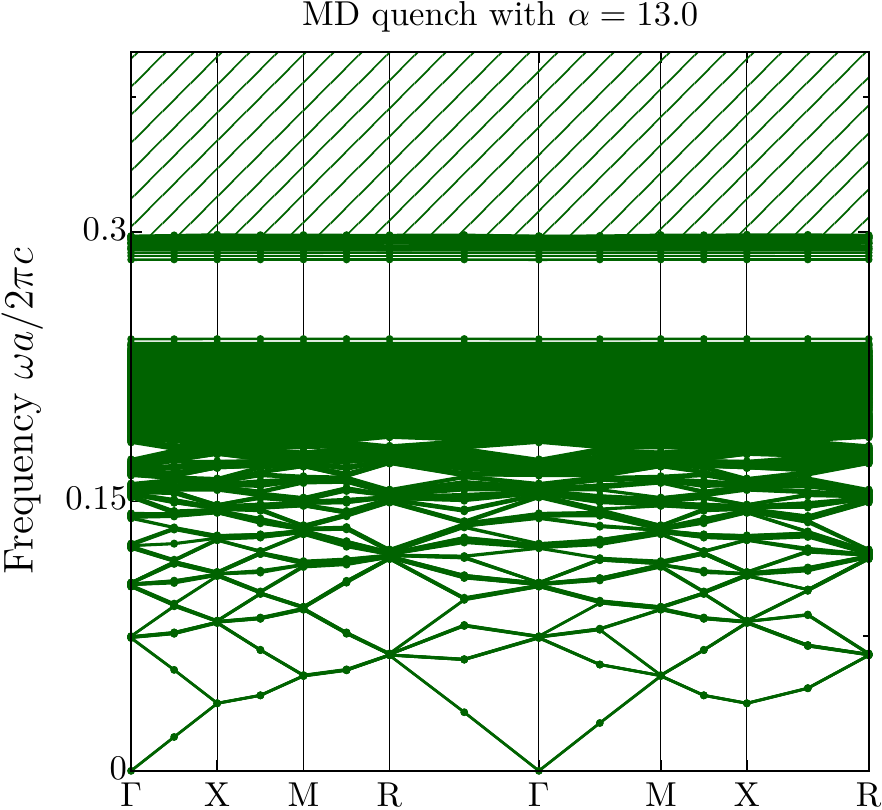}
  \hfill\hbox{}
  
  \hbox{}\hfill%
  \includegraphics[height=7.1cm]{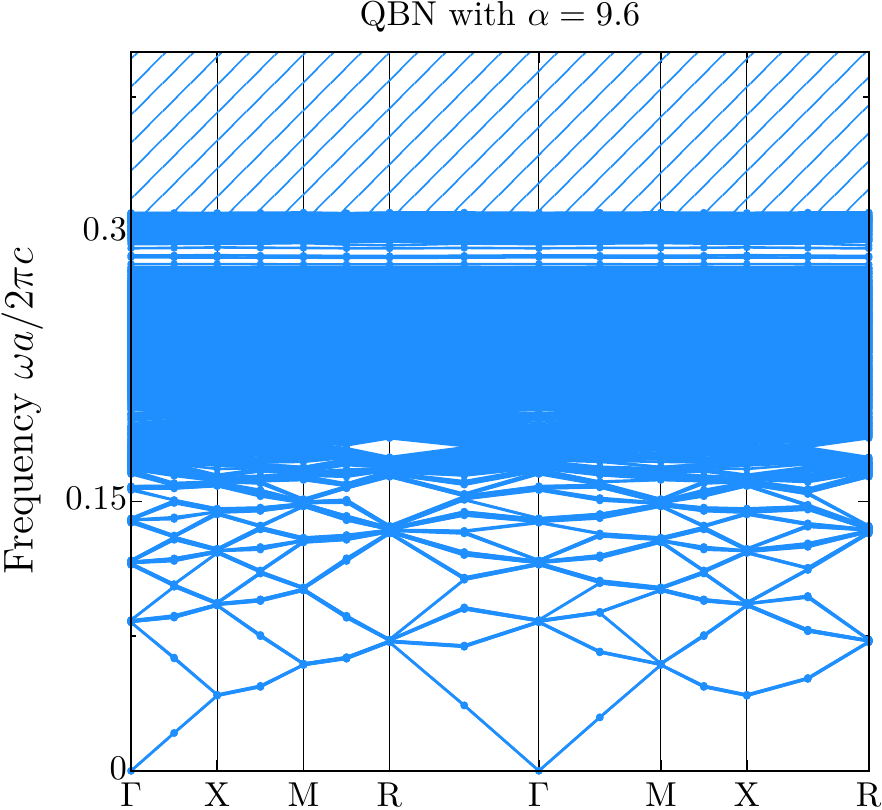}\hfill%
  \includegraphics[height=7.1cm]{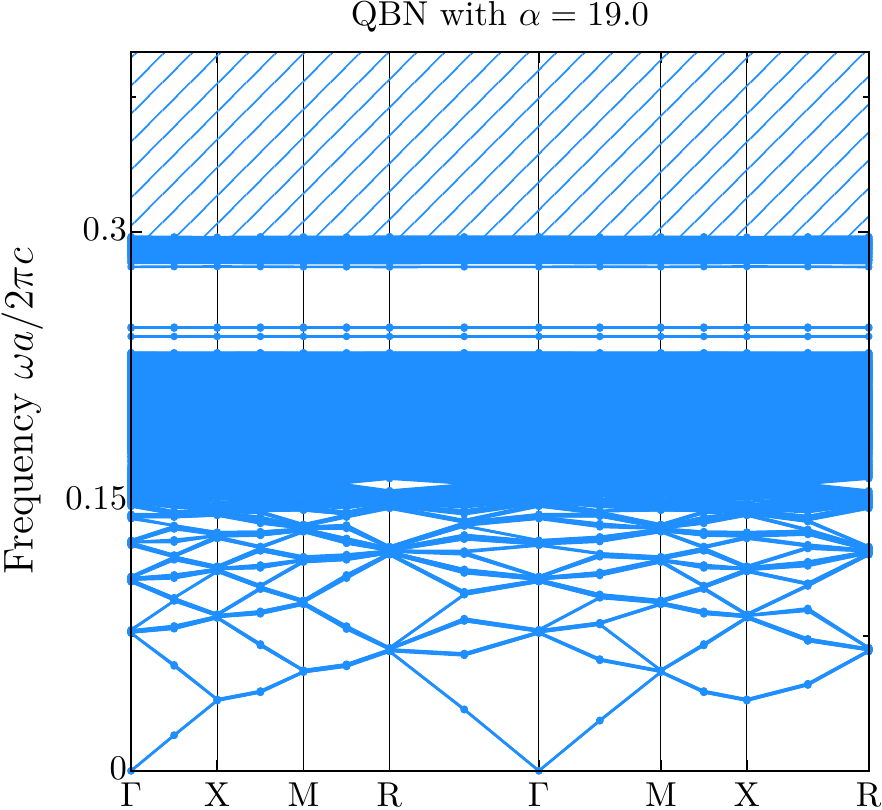}
  \hfill\hbox{}
  \caption{(Continued) Photonic band structures of (isotropic) disordered networks 
  at different dielectric contrasts $\alpha$; see also video~S2.}
  \label{fig:bst2}
\end{figure}

% #################################################################### %

\newpage
\subsection{S3.~Derivative of the optimal volume fraction}

\begin{figure}[h]
  \centering 
  \includegraphics[width=\linewidth]{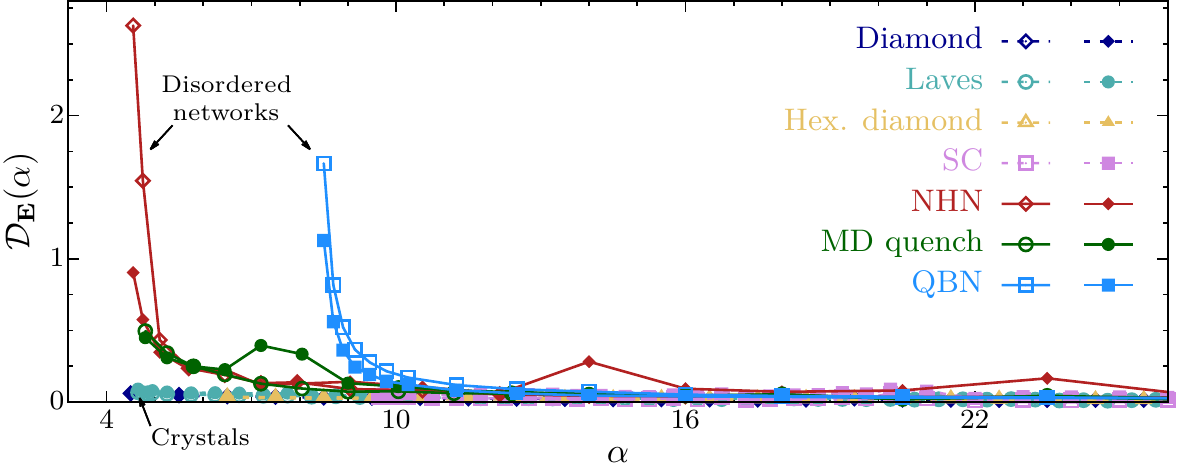}
  \caption{The average change of the electric field intensity 
  with the dielectric contrast $\alpha$ for the bands below (open 
  symbols) and above (solid symbols) the PBG.
  At low $\alpha$, the configuration of the electric field changes 
  more strongly for the disordered networks than for the photonic 
  crystals.
  \label{fig:delta-diff}}
\end{figure}

% #################################################################### %

%\newpage
\subsection{S4.~Derivative of the optimal volume fraction}

\begin{figure}[h]
  \centering 
  \includegraphics[width=\linewidth]{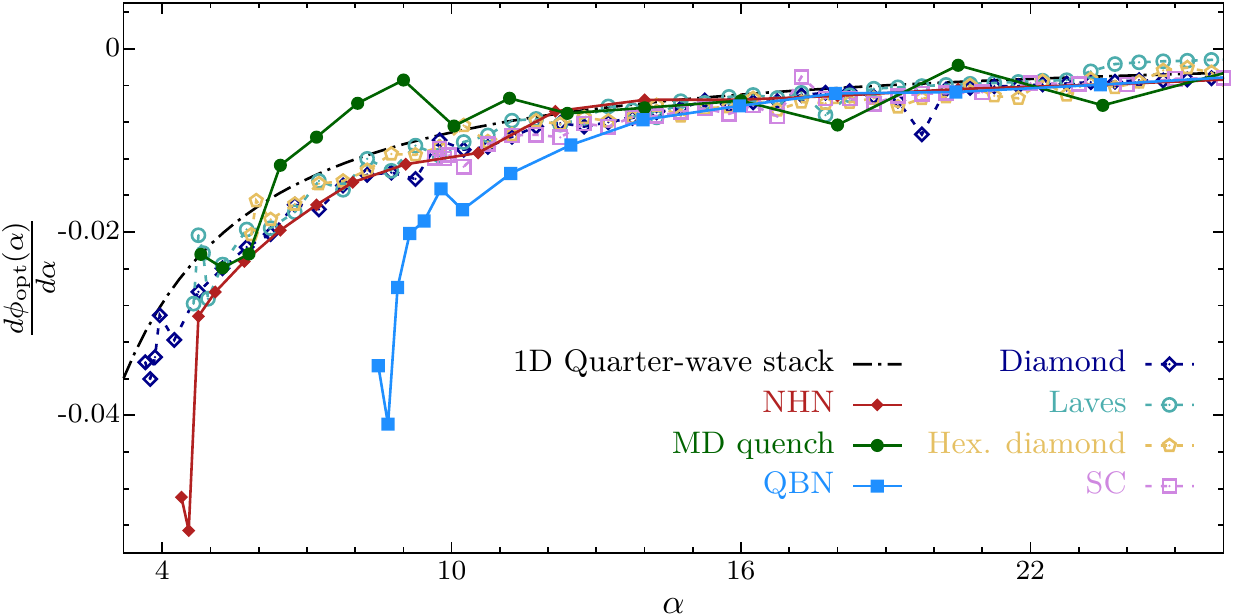}
  \caption{The derivative of the optimal volume fraction, 
  $d\phi_{\mathrm{opt}}(\alpha)/d\alpha$ for our crystal and disordered 
  networks approximately agrees at large $\alpha$ with the analytic 
  result for the 1D quarter-wave stack,
  $d\phi_{\QWS}(\alpha)/d\alpha=-1/(2\alpha^{3/2}+4\alpha+2\sqrt{\alpha})$,
  shown by the dashed-dotted line.
  \label{fig:filling-slope}}
\end{figure}

% #################################################################### %

\newpage
\subsection{S5.~System size effects}

\begin{figure}[h]
  \centering 
  \includegraphics[width=0.86\linewidth]{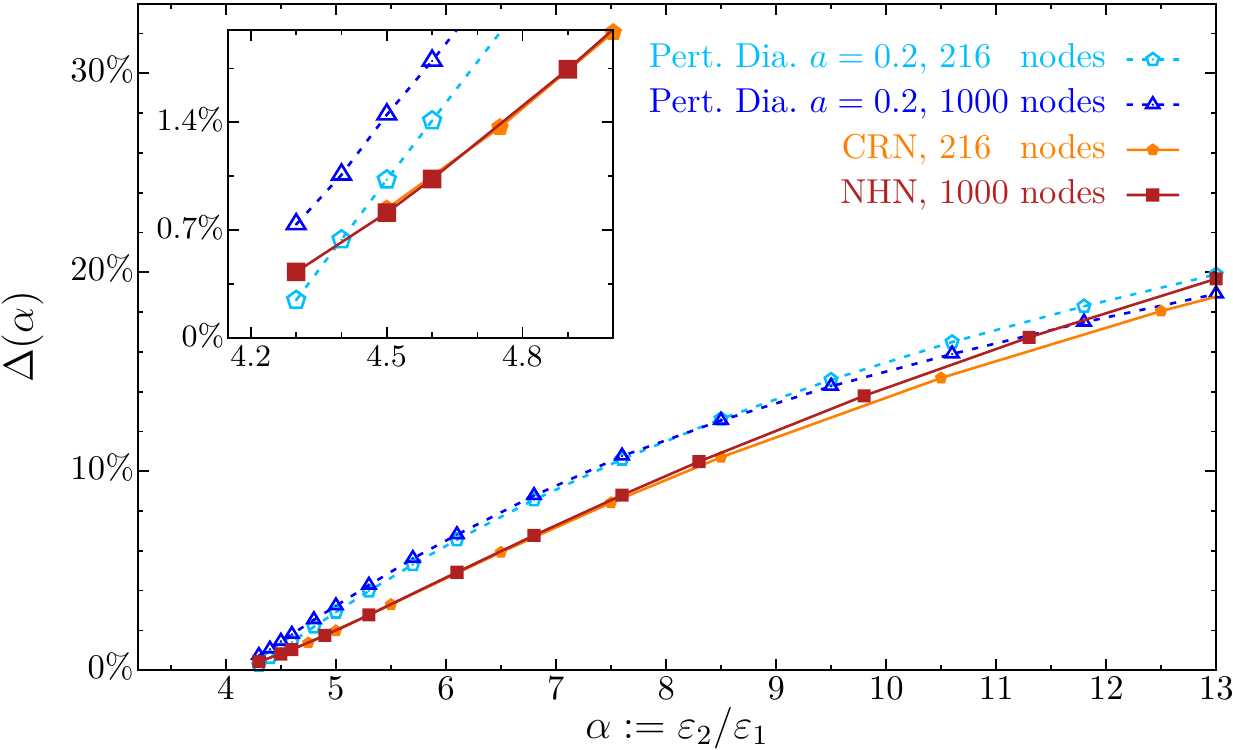}\\
  \includegraphics[width=0.86\linewidth]{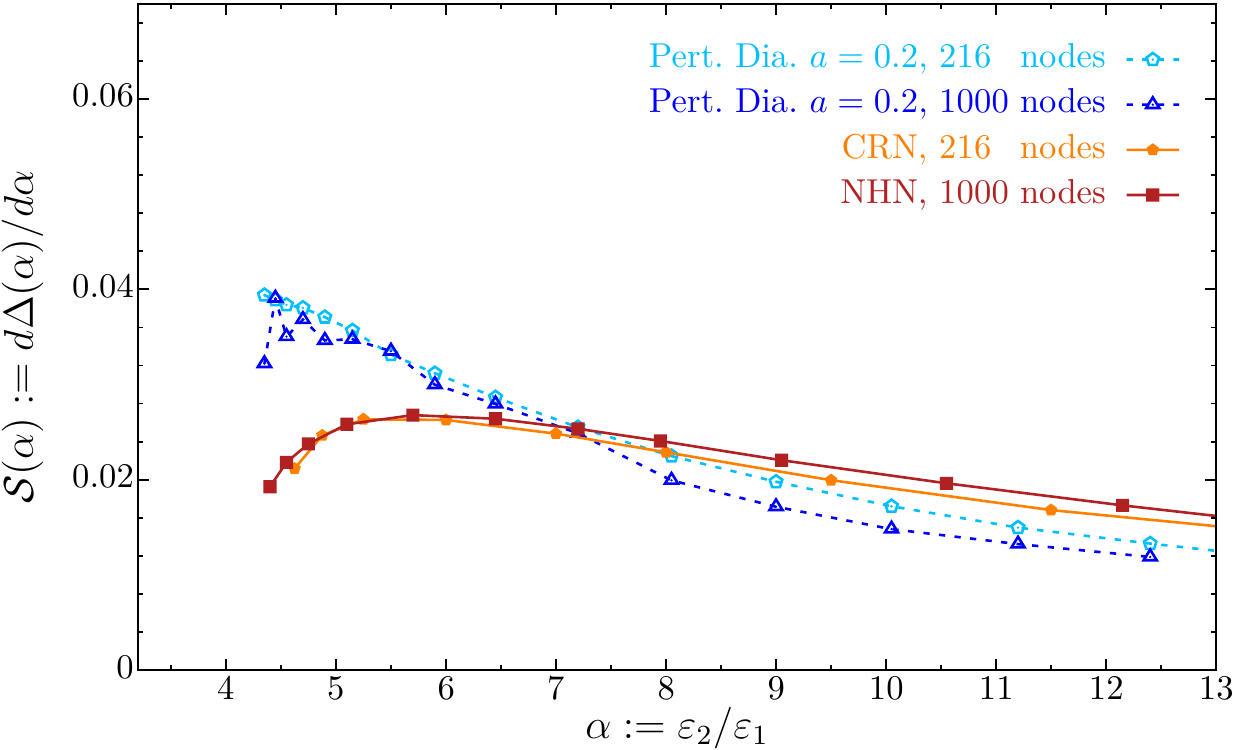}
  \caption{The \textit{gap plot} (top) and \textit{gap-sensitivity plot} 
  (bottom) for the NHN (1000 vertices) and the perturbed diamond network 
  with $a=0.2$ (216 vertices) are compared to similar networks at 
  different system sizes: a CRN with 216 vertices and a perturbed 
  diamond network with 1000 vertices, respectively.
  The curves at different system sizes agree within the systematic 
  errors and statistical fluctuations that can be expected between 
  different samples; that is, no strong system size effect is observed.
  \label{fig:sys-size}}
\end{figure}